\documentclass[sn-nature]{sn-jnl}
\usepackage{graphicx}%
\usepackage{multirow}%
\usepackage{amsmath,amssymb,amsfonts}%
\usepackage{amsthm}%
\usepackage{mathrsfs}%
\usepackage[title]{appendix}%
\usepackage{xcolor}%
\usepackage{textcomp}%
\usepackage{manyfoot}%
\usepackage{booktabs}%
\usepackage{algorithm}%
\usepackage{algorithmicx}%
\usepackage{algpseudocode}%
\usepackage{listings}%
\usepackage[separate-uncertainty=true,multi-part-units=single]{siunitx}
\usepackage{caption}
\usepackage{tabularx}
\usepackage[final]{pdfpages}

\DeclareSIUnit{\atomicunit}{au}

\newcommand{\lifetimeTheoretical}{\SI{41\pm 14}{\fs}}

\begin{document}

\title{Unexpected hydrogen dissociation in thymine: predictions from a novel coupled cluster theory}

\author[1,2,3]{\fnm{Eirik F.} \sur{Kjønstad}*}
\email{eirik.kjonstad@ntnu.no}

\author[1,2]{\fnm{O.~Jonathan} \sur{Fajen}}

\author[3]{\fnm{Alexander C.} \sur{Paul}}

\author[3]{\fnm{Sara} \sur{Angelico}}

\author[4]{\fnm{Dennis} \sur{Mayer}}

\author[4,5]{\fnm{Markus} \sur{Gühr}}

\author[1]{\fnm{Thomas J.~A.} \sur{Wolf}}

\author[1,2]{\fnm{Todd J.} \sur{Martínez}*}
\email{todd.martinez@stanford.edu}

\author[3]{\fnm{Henrik} \sur{Koch}*}
\email{henrik.koch@ntnu.no}

\affil[1]{\orgdiv{Department of Chemistry}, \orgname{Stanford University}, \orgaddress{\city{Stanford}, \state{CA}, \country{USA}}}

\affil[2]{\orgdiv{Stanford PULSE Institute}, \orgname{SLAC National Accelerator Laboratory}, \orgaddress{\city{Menlo Park}, \state{CA}, \country{USA}}}

\affil[3]{\orgdiv{Department of Chemistry}, \orgname{Norwegian University of Science and Technology}, \orgaddress{\city{Trondheim}, \postcode{7491}, \country{Norway}}}

\affil[4]{\orgname{Deutsches Elektronen-Synchrotron DESY}, \orgaddress{Hamburg}, \country{Germany}}

\affil[5]{\orgdiv{Institute of Physical Chemistry}, \orgname{University of Hamburg}, \orgaddress{Hamburg}, \country{Germany}}

\abstract{The fate of thymine upon excitation by ultraviolet radiation has been the subject of intense debate over the past three decades. 
Today, it is widely believed that its ultrafast excited state decay stems from a radiationless transition from the bright $\pi\pi^\ast$ state to a dark $n \pi^\ast$ state. 
However, conflicting theoretical predictions have made the experimental data difficult to interpret. Here we simulate the ultrafast dynamics in
thymine at the highest level of theory to date, performing wavepacket dynamics with a new coupled cluster method. 
Our simulation confirms an ultrafast $\pi\pi^\ast$ to $n \pi^\ast$ transition ($\tau = \lifetimeTheoretical$). 
Furthermore, the predicted oxygen-edge X-ray absorption spectra agree quantitatively with the experimental results. 
Our simulation also predicts an as-yet uncharacterized photochemical pathway: a $\pi\sigma^\ast$ channel that leads to hydrogen dissociation at one of the two N-H bonds in thymine. 
Similar behavior has been identified in other heteroaromatic compounds, including adenine, and several authors have speculated that a similar pathway may exist in thymine.
However, this was never confirmed theoretically or experimentally. This prediction calls for renewed efforts to experimentally identify or exclude the presence of this channel.}

\maketitle

\section*{Introduction}\label{sec:intro} 
Thymine, like other nucleobases, undergoes ultrafast radiationless relaxation back to the ground state after being excited by ultraviolet radiation. 
This property has been tied to the resilience of genetic material against light-induced damage \cite{crespo2004}.
However, the exact mechanism of this decay is not fully understood and has been a subject of debate for several decades. 
Experiments have identified at least two excited state decay channels, one with a lifetime on the order of $\lesssim$\SI{100}{\fs}, and one considerably longer, on the order of several ps \cite{canuel2005, ullrich2004, mcfarland2014, kang2002}.
Yet, the underlying mechanisms have been challenging to discern, with proposed explanations necessarily relying on simulations of the molecular dynamics. These simulations, in turn, introduce approximations with errors that are difficult to control. Different theoretical methods have therefore produced different explanations, and a consensus has yet to emerge.

Most reported simulations implicate two low-lying excited states in the relaxation: a dark $n\pi^\ast$ state (S$_1$) and the bright $\pi\pi^\ast$ state (S$_2$) into which the system is initially excited. 
Several simulations predict a $\pi\pi^\ast$ trapping channel in which the initial excitation to the $\pi\pi^\ast$ state is rapidly followed by relaxation into a minimum on the $\pi\pi^\ast$ surface, where the wavepacket is trapped for tens or hundreds of fs \citep{hudock2007,lischka2009, asturiol2009, asturiol2010exploring, barbatti2010, stojanovic2016new, MAI20179}. 
These simulations disagree, however, on the amount of $\pi\pi^\ast$ trapping, as well as the timescale and nature of the subsequent processes, with proposed mechanisms including $\pi\pi^\ast$ to $n\pi^\ast$ relaxation \citep{stojanovic2016new} and direct $\pi\pi^\ast$ relaxation to the ground state \citep{asturiol2009}. 
Some of these studies indicate that the amount of $\pi\pi^\ast$
trapping is reduced by improving the description of dynamical correlation (instantaneous electron-electron interactions) \citep{stojanovic2016new, asturiol2009}, a pattern that has also been found in the closely related nucleobase uracil \citep{chakraborty2021effect}. 
In line with this, an early density functional theory (DFT) study found significant $n\pi^\ast$ population within the first \SI{50}{\fs} \citep{picconi2011} and a more recent mixed-reference spin-flip DFT study also found rapid barrier-less $\pi\pi^\ast/n\pi^\ast$ transfer ($\tau = \SI{30}{\fs}$), including subsequent $n\pi^\ast$ trapping \citep{park2021impact}. 

Experimental evidence has implicated the $n\pi^\ast$ state in the early dynamics.
Indeed, by determining the gas phase oxygen-edge time-resolved X-ray absorption spectrum, Wolf et al.~\citep{wolf2017} found a fast component ($\tau = \SI{60 \pm 30}{\fs}$) which was attributed to population of the $n\pi^\ast$ state. Thus, the wavepacket appears to already transfer some of its population to the $n\pi^\ast$ state within the first \SI{100}{\fs}. 
Furthermore, Wolf et al.~found that the $n\pi^\ast$ signature lasts for several ps, revealing a second relaxation mechanism. 
After passing through the $\pi\pi^\ast/n\pi^\ast$ conical intersection, parts of the wavepacket appears to get trapped in a minimum on the $n\pi^\ast$ surface. 
While experiments have shown that the  $n\pi^\ast$ state is involved in the early sub-100 fs dynamics, it remains an open question whether or not there is some trapping in the $\pi\pi^\ast$ state \citep{wolfguehr2019, mayer2024time}. More accurate simulations of the dynamics are therefore essential to unravel the precise relaxation mechanism in thymine.

Here we present the highest-level wavepacket simulation on thymine to date. To the best of our knowledge, this is also the simulation with the highest level of electronic structure theory performed on a molecular system of this size. Thanks to recent developments \citep{kjonstad2017crossing, kjonstad2017resolving, kjonstad2019orbital, SchnackPetersen2022,kjonstad2023_cc_coupling,kjonstad2024scccoupling}, we were able to describe the electronic structure 
with the highly accurate coupled cluster singles and doubles (CCSD) \citep{purvis1982full} method.
This is the first time that this method is 
applied in nonadiabatic dynamics.
Coupled cluster (CC) theory is well-known for effectively capturing dynamical correlation, but it has been widely regarded as unsuited for excited state dynamics due to the presence of numerical artifacts at conical intersections \citep{hattig2005structure, kohn2007can, kjonstad2017crossing, Williams2023}.
Recent work has shown that these problems can be removed with similarity constrained CC (SCC) theory  \citep{kjonstad2017resolving,kjonstad2019orbital,kjonstad2024scccoupling}. 
Here, we apply the SCC with singles and doubles (SCCSD) method to perform molecular dynamics simulations.
To our knowledge, the present study is the first nonadiabatic dynamics simulation with a coupled cluster theory that describes conical intersections correctly. 
This demonstrates that the method is a viable electronic structure method for simulating photochemical phenomena.

\begin{figure}[H]
    \centering
    \includegraphics[width=0.85\linewidth]{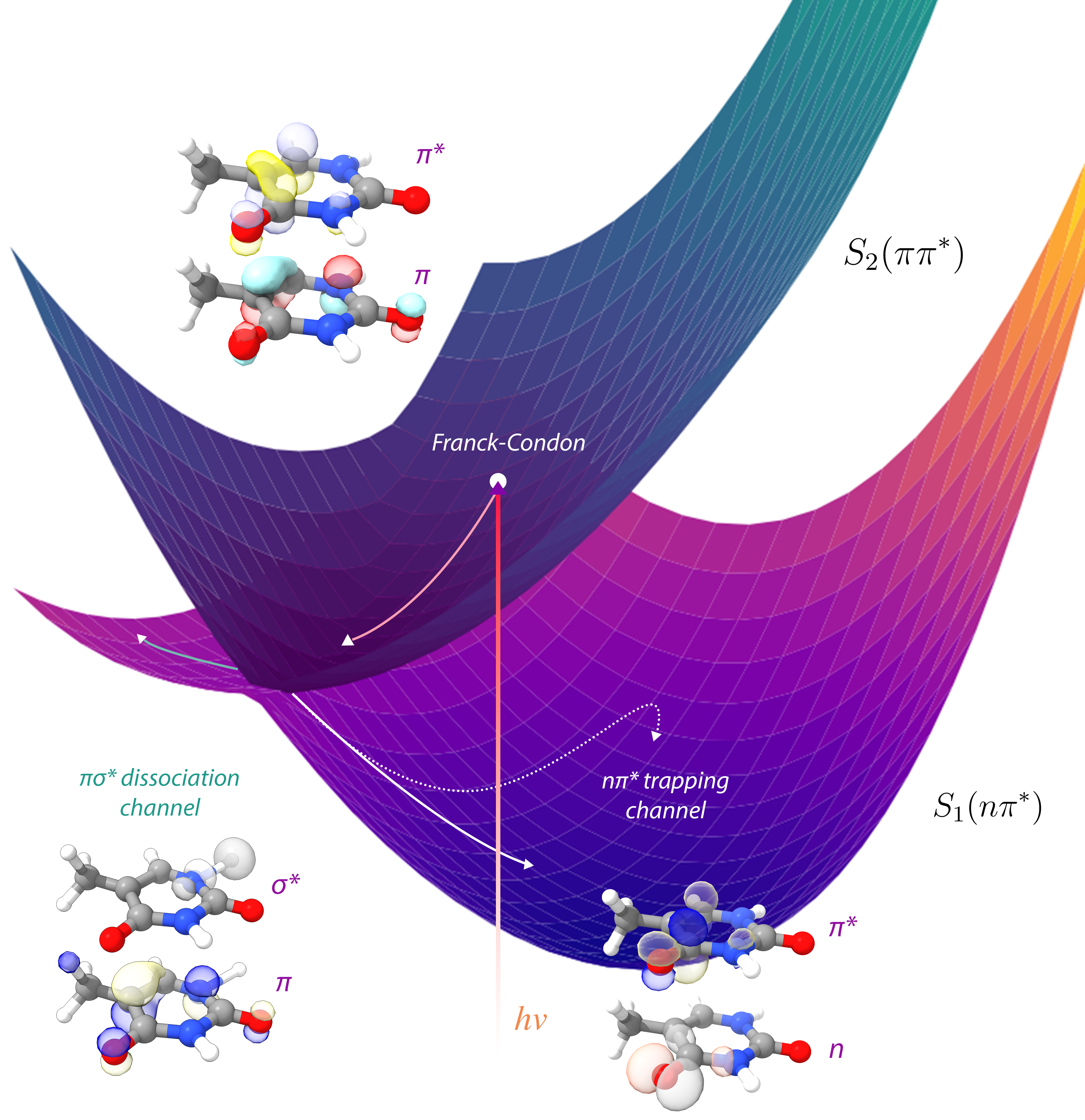}
    \caption{Photochemical pathways in the thymine simulation. 
    Following photoexcitation to the bright $S_2$($\pi\pi^\ast$) state, the simulation predicts two channels. The main channel (87\%) is the $n\pi^\ast$ trapping channel. Here, the wavepacket passes through the $S_1$/$S_2$ intersection and  heads toward a minimum on the $S_1(n\pi^\ast)$ surface. 
    This minimum is reached in two ways, either by heading to the minimum directly (solid line) or by reaching it indirectly through a $\pi\pi^\ast$ region on $S_1$ (dashed line). 
    The second channel (13\%) is an N-H dissociation channel. 
    Here, the wavepacket acquires $\pi\sigma^\ast$ character on $S_2$ (with the anti-bonding $\sigma^\ast$ orbital located at an N-H bond) before it passes through the $S_1$/$S_2$ intersection while retaining its $\pi\sigma^\ast$ character when it transfers to the $S_1$ surface. 
    Once the wavepacket is on $S_1$, N-H dissociation quickly follows as the wavepacket heads towards an intersection with the electronic ground state.}
    \label{fig:photochemical_pathways}
\end{figure}

\begin{figure}[H]
    \centering
    \includegraphics[width=\linewidth]{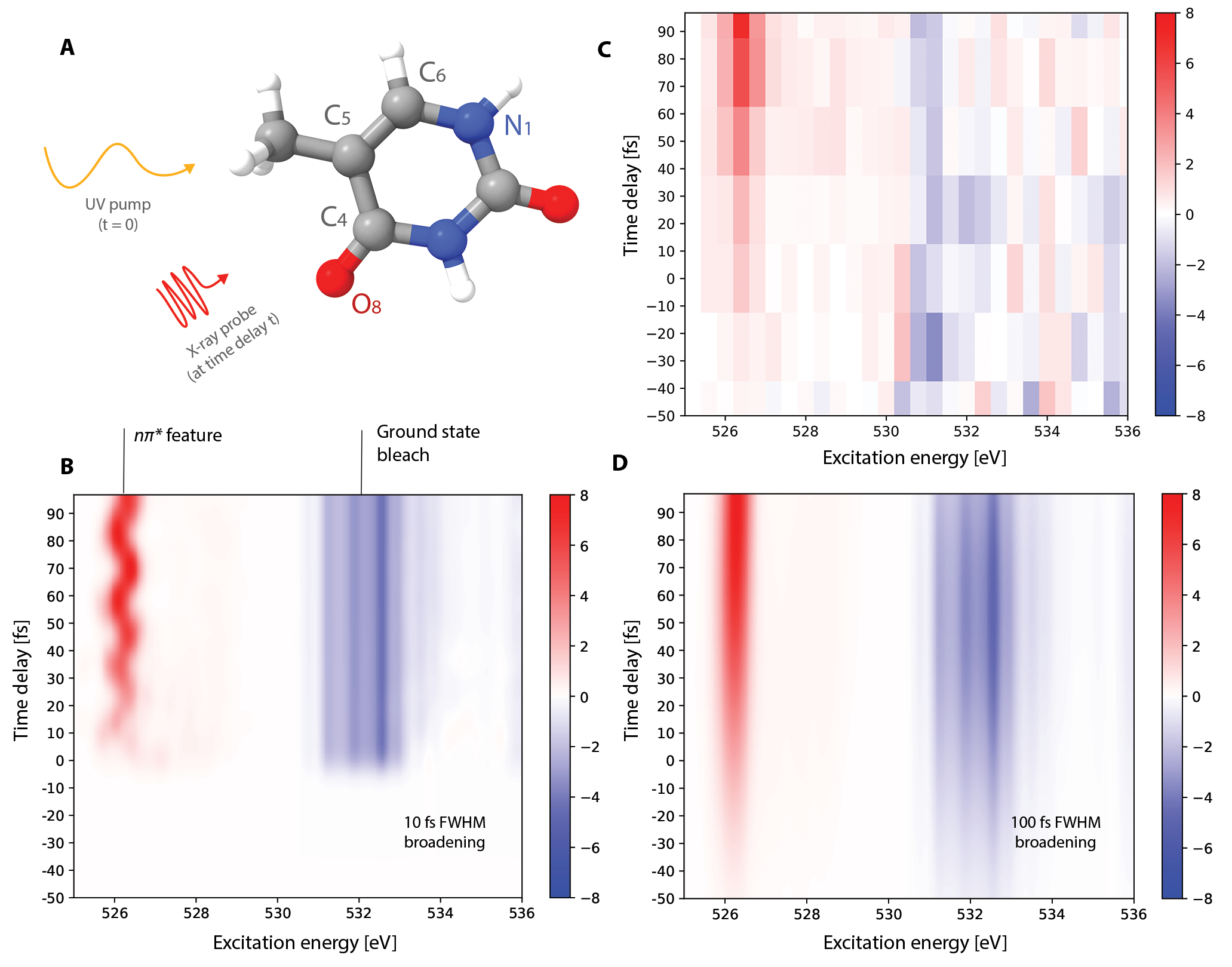}
    \caption{Thymine oxygen-edge X-ray absorption spectrum. 
    Panel A illustrates the pump-probe scheme for thymine, including the numbering used for the important atoms in the dynamics. 
    Panels B, C, and D show oxygen-edge X-ray absorption spectra for thymine, simulated (B and D) and experimental (C) \citep{wolf2017}. 
    Simulated spectra were computed using CC3/cc-pVDZ transition strengths and AIMS dynamics with CCSD/cc-pVDZ and SCCSD/cc-pVDZ, together with Gaussian broadening, \SI{0.3}{\eV}/\SI{10}{\fs} FWHM (panel B) and \SI{0.3}{\eV}/\SI{100}{\fs} FWHM (panel D). 
    For more details, see  Supporting Information S2. 
    The most prominent feature in the spectrum is a bright signal at \SI{526}{\eV}, which is due to the $n\pi^\ast$ state and reflects the rapid $\pi\pi^\ast$/$n\pi^\ast$ internal conversion after photoexcitation to the $\pi\pi^\ast$ state (at time $t = 0$). 
    The oscillations in the $n\pi^\ast$ feature (see panel B) are washed out by a larger time broadening (see panel D), reflecting the time resolution of the experiment (see panel C).}
    \label{fig:experimental_theoretical_spectrum}
\end{figure}

The main photochemical pathways in our simulation are illustrated in Figure \ref{fig:photochemical_pathways}. Our simulation confirms an ultrafast $\pi\pi^\ast$ to $n\pi^\ast$ conversion. 
In particular, the simulation predicts a time-resolved X-ray absorption spectrum (see Figure \ref{fig:experimental_theoretical_spectrum}) that captures the ultrafast component of the time-resolved oxygen-edge spectrum observed experimentally by Wolf et al.~\citep{wolf2017}. We show that this component is unequivocally associated with the $n\pi^\ast$ state, demonstrating that the ultrafast component of the decay is due to internal conversion from the $\pi\pi^\ast$ state. We further find no significant $\pi\pi^\ast$ trapping.

The interplay of a bright $\pi\pi^{*}$ and a dark $n\pi^{*}$ state is also central to the excited-state relaxation of a wide variety of other chromophores, including nucleobases such as uracil and adenine, although lifetimes and branching ratios differ. 
In adenine, another decay mechanism has been proposed, one in which a dark 
$\pi\sigma^{*}$ state is accessed through interconversion from the $\pi\pi^{*}$ state \cite{sobolewski2002,perun2005}. 
More generally, the importance of these low-lying $\pi\sigma^{*}$ states are well-recognized in a wide variety of heteroaromatic systems \citep{jourvet2004,jourvet2006,ashfold2006role,roberts2014role}. 
In adenine, the $\pi\sigma^{*}$ state is dissociative along the N-H bond coordinate and intersects with the ground state ($S_{0}$) at larger N-H distances. 
Following excitation to the $\pi\pi^{*}$ state, the wavepacket may reach this $\pi\pi^{*}/\pi\sigma^{*}$ conical intersection within \SI{20}{\fs}. 
Once on the repulsive $\pi\sigma^{*}$ state, N-H dissociation proceeds rapidly, resulting in a characteristically anisotropic total kinetic energy release (TKER) spectrum of H atom fragments \cite{nix2007}. 

Our high-level dynamics simulation predicts that thymine also has a similar $\pi\sigma^{*}$-mediated N-H dissociation channel (see Figure \ref{fig:photochemical_pathways}).
This finding comes as a surprise because no such dissociative channel has been previously reported.
In fact, TKER spectra across a broad range of wavelengths, 270--230 nm, showed no signature of ultrafast N-H dissociation \cite{schneider2006}, and recent experimental and theoretical investigations have not invoked this pathway to explain the molecular dynamics. 
Nonetheless, in analogy to the photochemistry of adenine and other heteroaromatics, it may be that certain regions of the $\pi\pi^{*}$ surface of thymine---in particular, regions accessed by a highly excited N-H stretching mode---lead to rapid relaxation through an intersection with a $\pi\sigma^{*}$ state, followed by hydrogen dissociation at the N-H bond.

\section*{Results and discussion}\label{sec:results_and_discussion}

Figure \ref{fig:photochemical_pathways} illustrates the two main photochemical pathways observed in the simulation. 
The first pathway is the $n\pi^\ast$ trapping channel (87\%). 
Here, the wavepacket travels directly from the Franck-Condon region to the $S_2$($\pi\pi^\ast$)/$S_1$($n\pi^\ast$) conical intersection, where it transfers population to the $S_1$($n\pi^\ast$) state. 
The transferred population then gets trapped in a minimum on the $S_1$($n\pi^\ast$) surface.
The second pathway is a dissociative $\pi \sigma^\ast$ channel (13\%). 
Here, the wavepacket acquires $\pi \sigma^\ast$ character on $S_2$, where the anti-bonding $\sigma^\ast$ orbital is localized at an N--H bond, leading to rapid hydrogen dissociation.

\begin{figure}[htb]
    \centering
    \includegraphics[width=\linewidth]{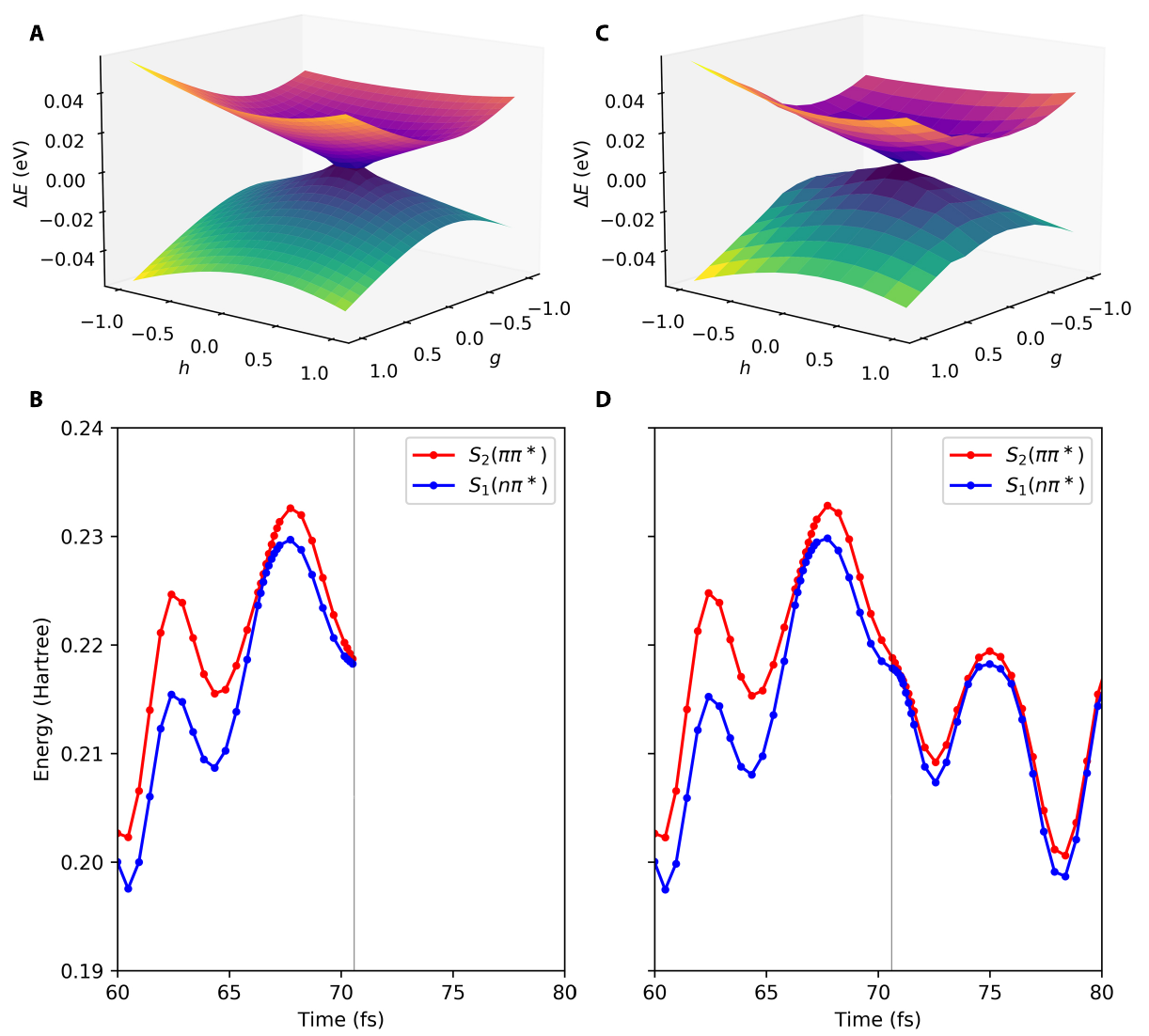}
    \caption{Dynamics with CCSD (panels A and B) and SCCSD (panels C and D). Panels A and C show a branching or $gh$ plane for an initial condition that encounters numerical artifacts in the simulations using CCSD, where $\Delta E$ is the energy difference between $S_1$ and $S_2$. 
    For CCSD (A), there is an ellipse of degeneracy in the $gh$ plane with unphysical complex-valued energies in the interior of the elliptical boundary; for SCCSD (C), we instead see a single point of degeneracy and no unphysical energies. 
    In CCSD simulations, the wavepacket may approach the intersection too closely and end up in the region with complex-valued energies (the interior of the ellipse in the $gh$ plane). 
    Whenever this happens, we re-run the simulation with SCCSD. 
    Panels B and D show the corresponding potential energy curves for the simulation with CCSD (B) and SCCSD (D). 
    At \SI{71}{\fs}, the CCSD simulation enters the complex-valued region and is terminated (B). 
    The SCCSD simulation, on the other hand, does not encounter any problems (D). 
    The conical intersection point used in the branching plane calculation is the geometry with the smallest $\Delta E$ (as given by SCCSD) in the nonadiabatic event at \SI{71}{\fs}.}
    \label{fig:ccsd_sccsd}
\end{figure}

The $n\pi^\ast$ trapping channel can be experimentally identified from the gas-phase time-resolved oxygen-edge X-ray absorption spectrum.
In Figure \ref{fig:experimental_theoretical_spectrum}, we compare the simulated and experimental \citep{wolf2017} X-ray absorption spectra. 
The most striking feature in these spectra is a bright signal, at around \SI{526}{\eV}, that grows in intensity in the first 50--\SI{70}{\fs}. 
The theoretical and experimental spectra are in agreement for this feature (see panels C and D), which we find to be due to population transfer from $S_2$($\pi\pi^\ast$) to $S_1$($n\pi^\ast$), in agreement with the mechanism proposed by Wolf et al.~\citep{wolf2017}. 
By analyzing the simulation data, we find a characteristic time of $\tau = \lifetimeTheoretical$ for this conversion (see Supporting Information S1), which fits well with experimental estimates ($\SI{60 \pm 30}{\fs}$ \citep{wolf2017} and $\SI{39 \pm 1}{\fs}$ \citep{Miura2023}). 
In the simulated spectrum, we also find weak but visible features that are associated with the initial $\pi\pi^\ast$ dynamics: in the first \SI{20}{\fs}, there are broad and diffuse features at around 526--528 eV and at \SI{534}{\eV}, where the \SI{534}{\eV} feature is partially hidden by the ground state bleach (see Figure 2B). 
These features reflect the rapid movement of the wavepacket away from the Franck-Condon region and towards the $S_2$($\pi\pi^\ast$)/$S_1$($n\pi^\ast$) conical intersection. 
Moreover, we find frequency oscillations associated with dynamics on the $n\pi^\ast$ state. These oscillations (at \SI{526}{\eV}) have a period of about \SI{20}{\fs} and an amplitude of about \SI{1}{\eV} (see panel B). 
The limited (\SI{100}{\fs} FWHM time resolution of the experiment \citep{wolf2017} washes these oscillations out (see Figure 2C), but we predict that they would be observed with improved time resolution. 
The oscillations are similarly washed out when we apply the same time broadening to the simulated spectrum (see Figure 2D).

An analysis of the simulation data, and in particular the growth of the $n\pi^\ast$ signal in the simulated spectrum, yields a $\pi\pi^\ast$/$n\pi^\ast$ conversion time of $\tau = \lifetimeTheoretical$.
This is consistent with the rate of $\pi\pi^\ast$/$n\pi^\ast$ conversion in the simulated dynamics, that is, from the observed change in character of the wavepacket from $\pi\pi^\ast$ to $n\pi^\ast$.
Analysis of the electronic character of the wavepacket reveals that its character does not always correspond to what one would expect from the population on $S_1$ and $S_2$, assuming that these correspond to $n\pi^\ast$ and $\pi\pi^\ast$, respectively. 
The adiabats mostly retain the orbital character in the Franck-Condon region, with $S_1$ mostly of $n\pi^\ast$ character and $S_2$ mostly of $\pi\pi^\ast$ character. 
However, the simulated spectrum shows that the adiabatic populations overestimate the true rate of internal conversion through the $\pi\pi^\ast$/$n\pi^\ast$ intersection; 
in particular, the adiabatic population transfer to $S_1$ is faster $(\tau = \SI{17 \pm 1}{\fs})$ than the growth of the $n\pi^\ast$ signal in the spectrum $(\tau = \SI{41 \pm 14}{\fs})$; see Supporting Information S1. 
This overestimation has two distinct causes. 
The first is that a part of the wavepacket that is transferred to $S_1$ retains its $\pi\pi^\ast$ character for some period of time, before eventually acquiring $n\pi^\ast$ character and making its way towards the $n\pi^\ast$ minimum. 
This slows down the appearance of the $n\pi^\ast$ signal compared to that expected from the adiabatic populations. 
The second is the $\pi\sigma^\ast$ character associated with the N$_1$-H dissociation channel. 
Here, the wavepacket that initially has $\pi\pi^\ast$ character acquires $\pi\sigma^\ast$ character on $S_2$ and retains it after conversion to $S_1$ (that is, $n\pi^\ast$ and $\pi\sigma^\ast$ flip energetic ordering). 
This conversion ($\pi\pi^\ast \rightarrow \pi \sigma^\ast$) contributes to the adiabatic $S_2 \rightarrow S_1$ population transfer but not to the $\pi\pi^\ast \rightarrow n\pi^\ast$ population transfer.
Indeed, when we reassign the populations according to the electronic character of the wavepacket, we find that the simulated spectrum is in close agreement with the diabatic populations $(\tau = 37 \pm 9$), thus showing that the \SI{526}{\eV} signal is due to the $n\pi^\ast$ character of the wavepacket. 

To better understand the dynamics, and in particular the $n\pi^\ast$ trapping channel, we identify stationary points on the $S_1$($n\pi^\ast$) and $S_2$($\pi\pi^\ast$) surfaces in the vicinity of the Franck-Condon region. 
In agreement with previous calculations, we find a minimum on the $S_1$($n\pi^\ast$) surface at an extended C$_4$--O$_8$ bond length \citep{stojanovic2016new,wolf2017}. 
This extension is due to the anti-bonding character of the $\pi^\ast$ orbital along the bond. We also find an $S_1$($n\pi^\ast$)/$S_2$($\pi\pi^\ast$) minimum energy conical intersection (MECI) that can be reached from the Franck-Condon region through C$_{5}$--C$_{6}$ elongation. 
Close to the intersection, CCSD exhibits numerical artifacts which can be removed with SCCSD. 
Figure \ref{fig:ccsd_sccsd} shows one of these artifacts as well as how it is corrected.

\begin{figure}[htb]
    \centering
    \includegraphics[width=\linewidth]{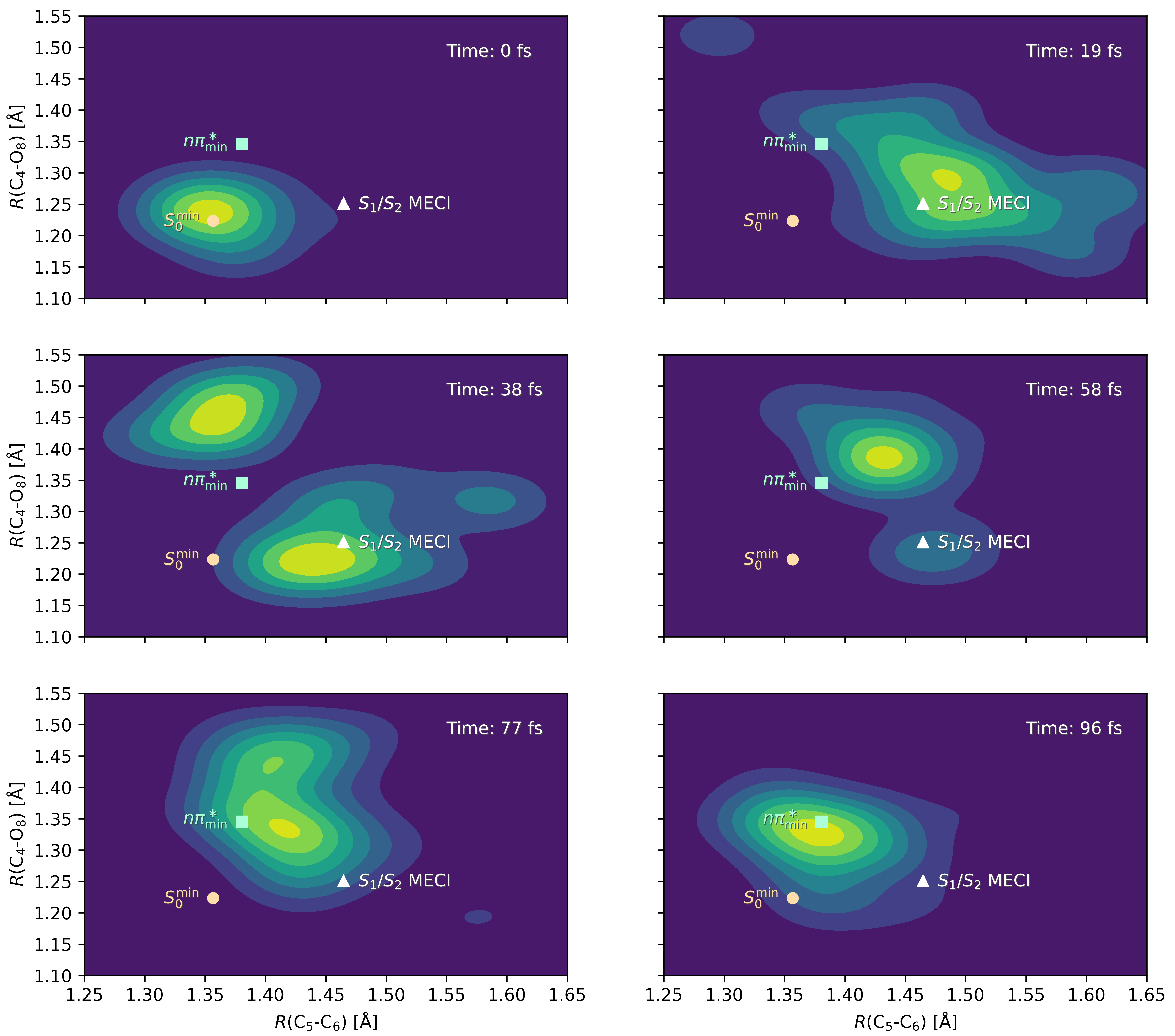}
    \caption{Nuclear density snapshots in the C$_5$-C$_6$ and C$_4$-O$_8$ coordinates. 
    Three important stationary points are shown: the ground state minimum $S_0^\mathrm{min}$, the minimum energy conical intersection $S_1/S_2$ MECI, and the $S_1$ minimum $n\pi^\ast_{\mathrm{min}}$. 
    The wavepacket quickly moves away from the Franck-Condon region ($S_0^\mathrm{min}$) and towards the minimum-energy conical intersection, where it starts transferring population to the $n\pi^\ast$ surface (\SI{19}{\fs}), eventually causing the wavepacket to split (\SI{38}{\fs}). 
    At longer times, after almost all of its population has transferred to the lower surface (\SI{58}{\fs}), the wavepacket settles in the vicinity of a minimum on the $n\pi^\ast$ surface (77 and \SI{96}{\fs}).}
    \label{fig:nuclear_density}
\end{figure}

By analyzing stationary points on $S_1$($n\pi^\ast$) and $S_2$($\pi\pi^\ast$),
Wolf et al.~\citep{wolf2017} suggested that the excited state decay of thymine follows a two-step process in the C$_5$-C$_6$ and C$_4$-O$_8$ coordinates (see Figure 2A for atom labelling): 
following photoexcitation, the C$_{5}$--C$_{6}$ bond is first elongated, and along this stretching coordinate, the $S_2$($\pi\pi^\ast$)/$S_1$($n\pi^\ast$) intersection seam is accessible; 
then, after interconversion to the $S_1$($n\pi^\ast$) state, the C$_4$--O$_8$ bond elongates as the wavepacket heads towards the $S_1$($n\pi^\ast$) minimum. This picture is borne out by our dynamics simulation. 
In Figure \ref{fig:nuclear_density}, we show the time evolution of the nuclear density in the C$_{5}$--C$_{6}$ and C$_4$--O$_8$ bond coordinates, and we indeed see this two-step process unfolding in real time.

\begin{figure}[H]
    \centering
    \includegraphics[width=\linewidth]{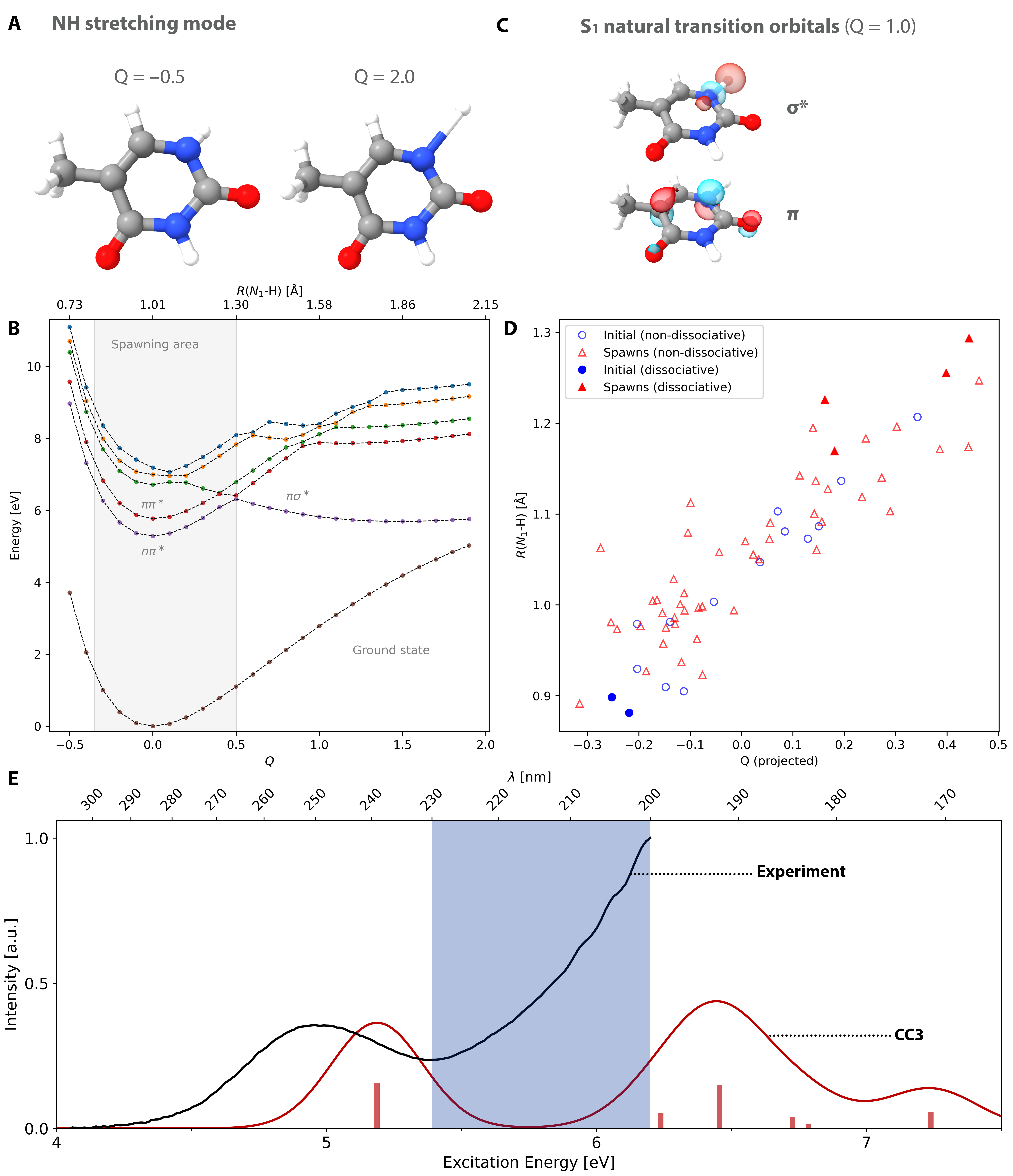}
    \caption{The $\pi\sigma^\ast$-mediated N-H dissociation pathway. 
    Panel A illustrates the normal mode associated with the N-H stretch, where $Q$ denotes the displacement along this mode relative to the ground state equilibrium geometry ($Q = 0$ corresponds to the $S_0$ minimum geometry). 
    Panel B shows potential energy curves along this normal mode, with the associated N-H bond length $R($N$_1$-H) given on the secondary axis. 
    The shaded area denotes the values of $Q$ for which we see spawning events in the dynamics. Panel C shows the natural transition orbitals of $S_1$ for $Q = 1.0$, where the state is of $\pi\sigma^\ast$ character. 
    Panel D shows the initial geometries (at time $t = 0$) and the spawning geometries (where population is transferred between $S_2$ and $S_1$), distinguishing between the dissociative geometries and the non-dissociative geometries. 
    For each geometry, we give the N-H bond length and the projected value of the mode displacement $Q$. Dissociating trajectories have negative initial values of $Q$ (short N-H distances) and spawn at high positive values of $Q$ (long N-H distances), corresponding to regions of the potential energy surfaces where the $\pi\sigma^\ast$ state becomes accessible. 
    Panel E shows the recorded ultraviolet gas-phase spectrum for thymine with a comparison to a simulated CC3 spectrum (see Supporting Information S9). 
    The shaded area corresponds to wavelengths where TKER spectra have not been recorded.}
    \label{fig:dissociation}
\end{figure}

The most intriguing result in the dynamics simulation is the presence of a  hydrogen dissociation channel (13\%). 
Figure \ref{fig:dissociation}  characterizes this channel in terms of the normal mode $Q$ associated with the N$_1$-H  stretch (see panel A). 
When thymine is displaced along $Q$ away from the equilibrium geometry ($Q = 0$), a dissociative $\pi\sigma^\ast$ state comes down and intersects with $n\pi^\ast$ and $\pi\pi^\ast$ states at around $Q = 0.4$ (see panel B), which corresponds to an N$_1$-H bond length of around $1.3$ Å. 
At longer N$_1$-H bond lengths, the character of $S_1$ is $\pi\sigma^\ast$ (see panel C). 
The two dissociating initial conditions both have a short initial N$_1$-H bond length, and they both transfer population to $S_1$ at long N$_1$-H bond lengths (see panel D). 
The simplest explanation for the dissociative channel, therefore, is that initial conditions with short N$_1$-H bonds experience a rapid extension of the bond in the initial phase of the nuclear dynamics, allowing them to access the $\pi\sigma^\ast$ state. 
This suggests that initial conditions with shorter N$_1$-H bonds at $t = 0$ would be more likely to dissociate. 
This indeed appears to be the case: 
out of 17 additional initial conditions specifically selected to have N$_1$-H bond lengths shorter than $0.9$ Å, we find that 7 of the conditions (41\%) have access to parts of the potential energy surfaces with $\pi\sigma^\ast$ character (see Supporting Information S6).

In the dynamics simulation, $2$ out of $16$ initial conditions (13\%) lead to hydrogen dissociation at the N$_1$-H bond.  
Inspection of the natural transition orbitals (NTOs) of one of the conditions shows how the dissociation happens (see Supporting Information S6). 
The other initial condition has similar behavior. 
Already at \SI{2.5}{\fs}, $S_{3}$ is mainly of $\pi\sigma^\ast$ character, while $S_{2}$ is still of $\pi\pi^\ast$ character and $S_{1}$ of $n\pi^\ast$ character. 
As the wavepacket moves along the N$_1$-H stretching coordinate, the $\pi\sigma^\ast$ state is stabilized, ultimately falling below the $\pi\pi^\ast$ and $n\pi^\ast$ states. 
The $S_{3}$/$S_{2}$ gap is about \SI{0.5}{\eV} at \SI{2.4}{\fs}, but by \SI{3.4}{\fs}, this gap has decreased to \SI{0.1}{\eV}. 
At the same time, $S_{3}$ is of mainly $\pi\sigma^\ast$ character at \SI{2.4}{\fs}, but by \SI{4.0}{\fs}, the character of $S_{2}$ and $S_3$ has flipped. 
Eventually, at around \SI{4.1}{\fs}, there is an intersection between $S_{1}$ and $S_{2}$, where most of the population is transferred to $S_{1}$. 
Moving away from the intersection, $S_{1}$ is dominated by $\pi\sigma^\ast$ character. 
{N$_1$-H} dissociation occurs rapidly over the next \SI{15}{\fs}, eventually reaching a ground state intersection at an extended N$_1$-H bond length (greater than 2.0 Å). 

Current experimental data does not allow us to verify or exclude the existence of the dissociative channel.
We do not find any clear signature of the $\pi\sigma^\ast$ channel in our simulated spectrum (see Figure \ref{fig:experimental_theoretical_spectrum}). 
However, the channel appears to be small ($\sim$13\%), and an oxygen-edge spectrum is not expected to be highly sensitive to changes  at the N$_1$-H bond. 
A nitrogen-edge spectrum, on the other hand, should be sensitive to these changes, as indicated by calculations at selected geometries (see Supporting Information S7). 
However, no experimental nitrogen-edge spectrum has been reported. 
Other experiments have appeared to disconfirm the existence of a dissociative $\pi\sigma^\ast$ channel.
Previous TKER spectra showed only smooth, isotropic H-atom kinetic energy, indicating no involvement of ultrafast, $\pi\sigma^\ast$-mediated N-H dissociation \cite{schneider2006}. 
However, these spectra only scanned over excitation wavelengths from 270--230 nm, and dissociation might become more common at shorter wavelengths.
For adenine, the TKER spectra for wavelengths from 280--234 nm also show smooth, isotropic H-atom kinetic energies. 
Only for wavelengths 233 nm and shorter do anisotropic, fast H-atom peaks consistent with ultrafast $\pi\sigma^\ast$-mediated N-H dissociation appear, and this signature becomes more intense with increasing excitation energy \cite{nix2007}. 
This behavior for adenine, coupled with the presence of the $\pi\sigma^\ast$ N-H dissociation in our dynamics, leads us to suggest that if the TKER experiments for thymine described by Schneider and coworkers \cite{schneider2006} are carried out using excitation wavelengths in the range 230--200 nm, then one might observe anisotropic, fast H-atom peaks consistent with this dissociative channel. 
However, wavelengths in the range 230--200 nm may not be directly comparable to our simulations as it would also excite the system to states above the $S_2(\pi\pi^\ast)$ state, in particular, to the $\pi\pi^\ast$ band that lies about 1.0 eV above $S_2(\pi\pi^\ast)$ (see Figure \ref{fig:dissociation}E and Supporting Information S9).
This is similar to adenine, where several states may contribute to the $\pi\sigma^\ast$ dissociation channel \citep{nix2007}. 
The utility of TKER experiments for confirming/denying the presence of the $\pi\sigma^\ast$ N-H dissociation channel in thymine has previously been suggested by others \cite{ashfold2010,roberts2014role}. 

\clearpage

The ultimate fate of the dissociative $\pi\sigma^\ast$ channel cannot be resolved in this work, as CC theory is known to break down when the $\pi\sigma^\ast$ state becomes degenerate with the ground state (the dissociation limit). 
Because of this, there may also be a pathway back to the ground state through an intersection with the ground state, which might compete with N-H bond fission \citep{roberts2014role}. However, the presence and accessibility of the $\pi\sigma^\ast$ channel is well-described in our simulations. Both
the early-time involvement of the $\pi\sigma^\ast$ state, and the initial rapid N-H stretch, are treated correctly with CC theory, and this aspect of the simulated dynamics paints a picture in line with both theoretical and experimental studies of $\pi\sigma^\ast$ states in other heteroaromatics \citep{ashfold2006role, roberts2014role}. 

\section*{Conclusion}\label{sec:conclusion}
The photorelaxation pathways of thymine are still under debate and a clear consensus has yet to emerge, despite numerous theoretical and experimental investigations.
Here, we have simulated the ultrafast dynamics of thymine using high-level nonadiabatic dynamics simulations (\emph{ab initio} multiple spawning) combined with coupled cluster theory (with single and double excitations) for the electronic structure. 
This represents, to the best of our knowledge, the highest level simulation performed on thymine to date, and the simulation with the highest level of electronic structure on a molecular system of this size. From the data, we simulated the oxygen-edge X-ray absorption spectrum of Wolf et al. \citep{wolf2017}, obtaining excellent agreement on the appearance and position of a signal unequivocally associated with the $n\pi^\ast$ state, as well as for the associated $\pi\pi^\ast/n\pi^\ast$ conversion time: 
our theoretical estimate of $\lifetimeTheoretical$ agrees quantitatively with experimental estimates ($\SI{60 \pm 30}{\fs}$ and {$\SI{39 \pm 1}{\fs}$} \citep{wolf2017,Miura2023}). 
Furthermore, we find no significant $\pi\pi^\ast$ trapping.

Interestingly, our simulation predicts an additional, minor channel in which population is rapidly transferred to a $\pi\sigma^\ast$ state, leading to N-H dissociation. 
This type of channel has been implicated in the ultrafast excited state dynamics of other heteroaromatic systems, including adenine \citep{ashfold2006role,roberts2014role}. 
In view of this surprising finding, we believe further experiments---for example, measuring the nitrogen-edge X-ray absorption spectrum---are warranted to confirm or disprove the existence of this dissociative channel as one of the relaxation pathways in thymine. 

This work was made possible by  developments in coupled cluster theory. 
It is the first nonadiabatic dynamics simulation based on this electronic structure method (CCSD), and, furthermore, the first simulation based a coupled cluster method that correctly describes conical intersections. 
Earlier work by some of the authors \citep{kjonstad2017crossing,kjonstad2017resolving,kjonstad2019orbital} had already indicated that the method could be modified to correctly describe conical intersections and therefore also nonadiabatic dynamics. 
Here, we have shown that the method introduced in these papers, the similarity constrained coupled cluster method, can in fact be applied in nonadiabatic dynamics simulations, opening up a range of applications that may now be studied with the hierarchy of coupled cluster methods. 
Given their ability to capture dynamical correlation, we expect that these methods will shed new light on the photochemistry of a variety of systems.

\section*{Methods}\label{sec:meThods}
\subsection*{Theoretical}
The excited state dynamics simulation was performed with the \emph{ab initio} multiple spawning (AIMS) method \citep{ben2000ab, ben2002ab,curchod_aims_chemrev_2018}. 
We prepared 16 initial conditions (ICs) and simulated the dynamics for a total of \SI{4000}{\atomicunit} ($\SI{\sim100}{\fs}$). 
The initial trajectory basis functions (TBFs) were sampled from a \SI{0}{\K} harmonic Wigner distribution obtained from the ground state equilibrium geometry at the CCSD/aug-cc-pVDZ level. 
Both the geometry and the frequencies were obtained with CCSD/aug-cc-pVDZ. The dynamics simulation was performed at the CCSD/cc-pVDZ level. 
For the ICs, we adopt the independent first generation approximation, that is, we average results over 16 independent AIMS simulations. 
Of these, 12 run normally for the whole simulation. 
Two ICs reach intersections between $S_{0}$ and $S_{1}$ around \SI{20}{\fs} and are terminated at this point because CCSD cannot describe $S_{0}$/$S_{1}$ intersections (termination is caused by convergence problems). 
A further two ICs enter a defective region surrounding the $S_{1}$/$S_{2}$ intersection, at which point the energies become complex-valued and the simulations are terminated. 
These two ICs were rerun with similarity constrained CC theory (SCCSD) using the $\mathscr{E}$ with $T = 0$ projection \citep{kjonstad2024scccoupling}. 
For both CCSD and SCCSD, the coupling elements were evaluated with the nuclear derivative acting on the right vector without normalization; 
for more details, see Refs.~\citenum{kjonstad2023_cc_coupling,kjonstad2024scccoupling}. 
In the time-integration, we have used a default timestep of \SI{20}{\atomicunit} (\SI{0.5}{\fs}) and a smaller timestep of \SI{5}{\atomicunit} (\SI{0.1}{\fs}) in regions  of high coupling. 
The spawning threshold was set to 0.05, where the spawning criterion is given as the norm of the coupling times the velocity. 
This produced a total of 67 TBFs over the course of the simulation.

Using the data from the dynamics simulation, we simulated the oxygen-edge X-ray absorption spectrum using the CC with approximate triples (CC3) \citep{koch1997cc3} method with the cc-pVDZ basis. 
Core excited states were obtained with the core-valence separation approximation \citep{coriani2015communication}. 
To simulate the spectrum, we have applied the incoherent approximation \citep{list2020probing}, where the spectrum is calculated as an average of the spectra computed at the centers of the TBFs, weighted with the corresponding TBF amplitude. 
To avoid state assignment ambiguities, we have used CCSD for the valence excited states and CC3 for the core excited states, where the strengths are evaluated with CCSD (excluding the approximate triples in the CC3 states). 
Similarly, absorption energies are evaluated from CCSD-to-CC3 energy differences. 
All X-ray absorption spectra were shifted by \SI{-0.5}{\eV}, corresponding to the required shift needed to align the first ground state peak with the experimental value at \SI{531.4}{\eV}. 
The UV-vis absorption spectrum was calculated at the CC3/aug-cc-pVDZ+KBJ(3-4) level for the ground state equilibrium geometry, 
determined with CCSD/aug-cc-pVDZ, see Supporting Information S9.

All electronic structure calculations were performed using development versions of the eT program \citep{eT2020}, version 1.8 for the dynamics and version 1.7 for the spectra. 
The AIMS dynamics was run with the FMS program and an interface to the eT program. 
See Supporting Information S8 for a description of the interface.

\subsection*{Experimental}
The experimental UV spectrum of thymine was taken with a Cary 5E UV-Vis-NIR spectrometer using an in-house developed gas cell described in Ref.~\cite{Mayer2022CP}. 
The sample was purchased from Sigma-Aldrich with >99\% purity and used without further refinement. 
The cell was heated up to \SI{150}{\degreeCelsius} to obtain sufficient absorption. Spectra were recorded over a range of 200--\SI{400}{\nm} with a step size of \SI{0.5}{\nm} and an integration time of \SI{0.5}{\second} per data point. 
A background spectrum of the empty cell has been recorded at the same temperature and settings and subtracted from the spectra including the sample.

\backmatter

\bmhead{Acknowledgments}
This work was supported by the Norwegian Research Council through FRINATEK project 275506, the European Research Council (ERC)
under the European Union’s Horizon 2020 Research and Innovation Program
(Grant No.~101020016), and the AMOS program within the U.S. Department of Energy (DOE), Office of Science, Basic Energy Sciences, Chemical Sciences, Geosciences, and Biosciences Division. 
OJF is a U.S. Department of Energy Computational Science Graduate Fellow (Grant No.~DE-SC0023112). 
We acknowledge computing resources through UNINETT Sigma2--the National Infrastructure for High Performance Computing and Data Storage in Norway, project NN2962k.

\bibliography{sn-bibliography}

\clearpage


\section*{Supplementary material}

# Supporting information for "Unexpected hydrogen dissociation in thymine: predictions from a novel coupled cluster theory"

Eirik F. Kjønstad*[1,2,3], O. Jonathan Fajen[1,2], Alexander C. Paul[3], Sara Angelico[3], Dennis Mayer[4], Markus Gühr[4,5], Thomas J. A. Wolf[1], Todd J. Martínez*[1,2], Henrik Koch*[3]

[1]Department of Chemistry, Stanford University, Stanford, CA, USA.
[2]Stanford PULSE Institute, SLAC National Accelerator Laboratory, Menlo Park, CA, USA.
[3]Department of Chemistry, Norwegian University of Science and Technology, Trondheim, 7491, Norway.
[4]Deutsches Elektronen-Synchrotron DESY, Hamburg, Germany.
[5]Institute of Physical Chemistry, University of Hamburg, Hamburg, Germany.

Contributing authors: eirik.kjonstad@ntnu.no; todd.martinez@stanford.edu; henrik.koch@ntnu.no;

## S1 Estimation of time constants: the rate of $\pi\pi^*$ decay and of $\pi\pi^*/n\pi^*$ internal conversion

In order to model the kinetics of $\pi\pi^*$ decay and the $\pi\pi^*/n\pi^*$ internal conversion, we first assume that the $\pi\pi^*$ state is populated at time $t = 0$, that is, $P_{\pi\pi^*}(0) = 1$ and $P_{n\pi^*}(0) = 0$. We further assume a partial transfer of population from $\pi\pi^*$ to $n\pi^*$ that decays exponentially. In the simulation, all initial conditions begin on the $\pi\pi^*$ state and the majority of them become trapped in the $n\pi^*$ state, consistent with the above two assumptions. In more detail, we have

$$P_{\pi\pi^*}(t) = 1 - C_2(1 - \exp(-t/\tau_{\pi\pi^*})) \qquad (1)$$
$$P_{n\pi^*}(t) = C_1(1 - \exp(-t/\tau_{n\pi^*})). \qquad (2)$$

The time constants $\tau_{\pi\pi^*}$ and $\tau_{n\pi^*}$ give the characteristic time for $\pi\pi^*$ decay and the rate of $\pi\pi^*/n\pi^*$ conversion, respectively.

This simple model does not give a complete picture of the observed dynamics. In the simulation, we find two initial conditions that terminate at an $S_1/S_0$ intersection. Hence, there is a second decay channel (in particular, $\pi\pi^*$ to $\pi\sigma^*$). The proportion of dissociative decay is assumed to be sufficiently small to be neglected in the models. This should be a good approximation since only a small number of conditions participate in the channel (2/16). In the population statistics below, we assume that these two conditions decay instantly to $S_0$ when they are terminated.

## S1.1 Adiabatic populations and time constants

An initial estimate of the $\pi\pi^*/n\pi^*$ internal conversion time constant $\tau_{n\pi^*}$ and the $\pi\pi^*$ lifetime $\tau_{\pi\pi^*}$ can be extracted directly from the adiabatic populations. We here assume that $S_1$ is always of $n\pi^*$ character, and $S_2$ of $\pi\pi^*$ character, that is,

$$P_{S_2}(t) = P_{\pi\pi^*}(t) \tag{3}$$
$$P_{S_1}(t) = P_{n\pi^*}(t) \tag{4}$$

The adiabatic populations, along with time constants, are shown in Figure S1. In terms of the adiabats, we find $\tau_{\pi\pi^*} = 18 \pm 1\,\text{fs}$ and $\tau_{n\pi^*} = 17 \pm 1\,\text{fs}$. These estimated time constants are too low, as we can show by considering the character of the wavepacket.

## S1.2 Diabatic populations and time constants

To find more accurate estimates, we can reassign the populations according to the state character of the adiabatic states, thereby directly estimating $P_{\pi\pi^*}(t)$ and $P_{n\pi^*}(t)$. To calculate the contribution from a populated adiabatic state to the constituent diabatic states, $\pi\pi^*$ and $n\pi^*$, we assume that the electronic transitions from the ground state to each of the diabats remain well-defined as bright and dark, respectively. Given that the deviations from planarity are generally small, this should be a good approximation. For a populated adiabatic state, $S_k$, we assign the population to $\pi\pi^*$ and $n\pi^*$ according to the oscillator strengths $f_{if}$ ($i$ = initial, $f$ = final). In particular, if $k = 2$ (normally $\pi\pi^*$), then

$$P^k_{\pi\pi^*} = \frac{\max(f_{01}, f_{02})}{f_{\text{tot}}}, \quad P^k_{n\pi^*} = 1 - P^k_{\pi\pi^*}, \tag{5}$$

where $f_{\text{tot}} = f_{01} + f_{02}$, and if $k = 1$ (normally $n\pi^*$), then

$$P^k_{n\pi^*} = 1 - \frac{\min(f_{01}, f_{02})}{f_{\text{tot}}}, \quad P^k_{\pi\pi^*} = 1 - P^k_{n\pi^*}. \tag{6}$$

This procedure usually results in a clear assignment, with one adiabatic state accounting for $\geq 99\%$ of the total brightness. In the vicinity of the $S_1(n\pi^*)/S_2(\pi\pi^*)$ intersection, however, $S_1$ and $S_2$ assume more mixed electronic characters, and the

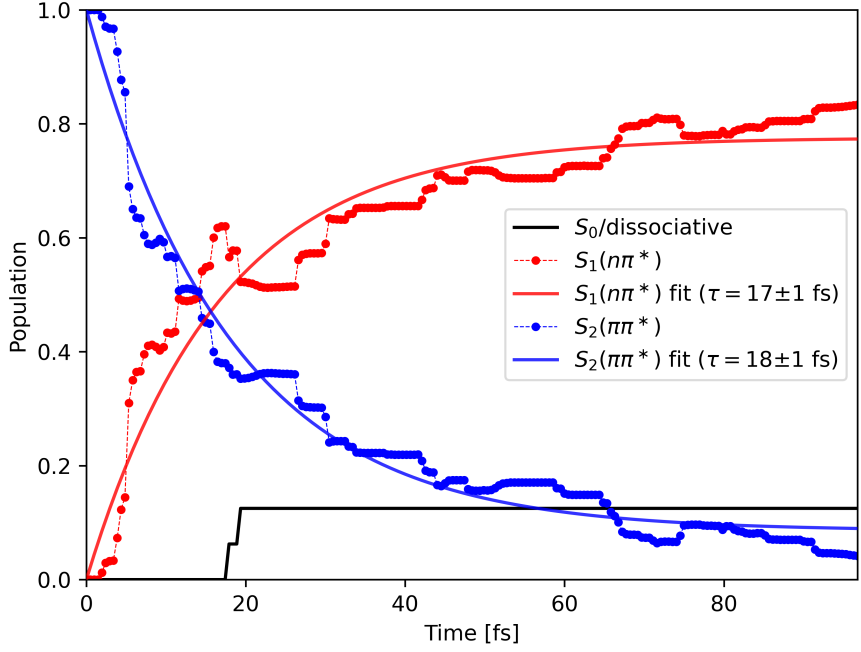

**Fig. S1**: Adiabatic poopulations for $S_1$ and $S_2$ with associated lifetimes. The error bars on the characteristic time $\tau$ is reported as two standard deviations. The two dissociative initial conditions terminate at around 20 fs and are here taken to transfer completely to $S_0$ through the dissociative channel.

reassigned population is more evenly shared between the diabats. Calculated diabatic populations, along with estimated time constants, are shown in Figure S2. From the diabatic populations, we find that the time constants for conversion are

$$\tau_{\pi\pi^*} = 34 \pm 6 \,\text{fs} \tag{7}$$
$$\tau_{n\pi^*} = 37 \pm 9 \,\text{fs}, \tag{8}$$

where error bars are given as two standard deviations. Clearly, the $\pi\pi^*$ to $n\pi^*$ transfer is slower than expected from the adiabats. Notice that at short and long times, the adiabatic and diabatic populations agree, indicating a clear separation of $n\pi^*$ and $\pi\pi^*$ character close to the Franck-Condon point and close to the $n\pi^*$ minimum. At intermediate times (around 10 fs to 50 fs), state-mixing prevents a clear, binary assignment of the adiabats. Furthermore, although transfer to $S_1$ occurs rapidly from 10 fs to 30 fs, this is not accompanied by a similar change from $\pi\pi^*$ to $n\pi^*$ character. The less rapid population transfer of the diabats, as compared to the adiabats, is consistent with the slower appearance of the $n\pi^*$ signal in the simulated spectrum, as can be seen from Figure 2. This fact is discussed further in Section S1.3.

Indeed, there are multiple instances of population transfer from $S_2$ to $S_1$ in which the electronic state retains $\pi\pi^*$ character for 10-15 fs. These $S_1/\pi\pi^*$ trajectories demonstrate unique behavior in the $C_5$–$C_6$ and $C_4$–$O_8$ plane, as seen in Figure S3. The trajectories on $S_1$ with $\pi\pi^*$ character move toward longer $C_5$–$C_6$ and $C_4$–$O_8$ distances before turning around and moving toward the much shorter $C_5$–$C_6$ distance and longer $C_4$–$O_8$ characterizing the $n\pi^*$ minimum, acquiring $n\pi^*$ character along the way. Trajectories that spawn on $S_1$ with $n\pi^*$ character, on the other hand, move directly toward longer $C_4$–$O_8$ and shorter $C_5$–$C_6$ distances.

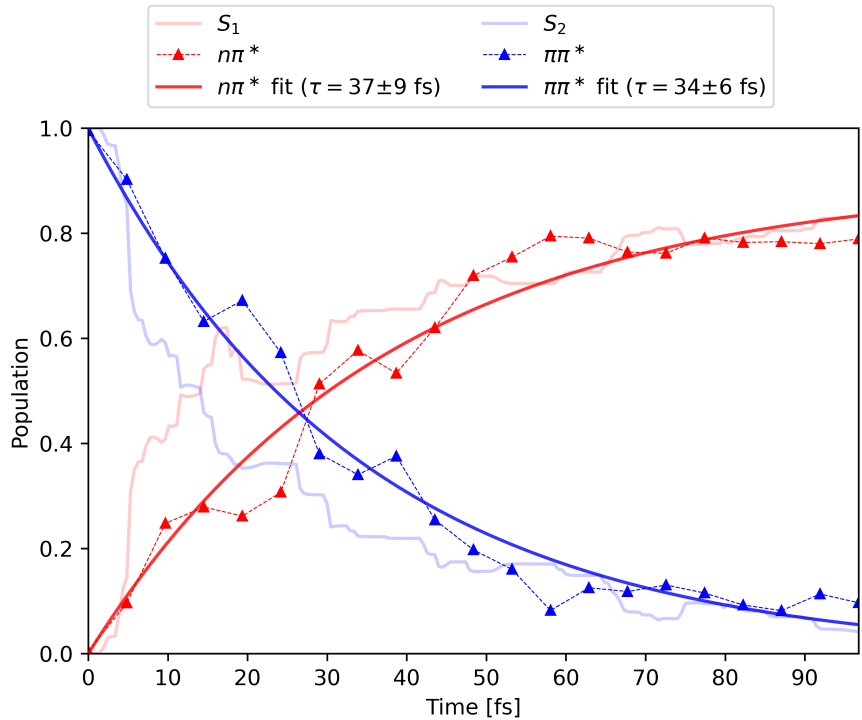

**Fig. S2**: Diabatic populations for $n\pi^*$ and $\pi\pi^*$ with associated lifetimes. The error bars on the characteristic time $\tau$ is reported as two standard deviations. The two dissociative initial conditions terminate at around 20 fs and are here taken to transfer completely to $S_0$ through the dissociative channel. The population in this channel is not shown. The adiabatic populations are shown in solid lines and illustrate the more rapid transfer of the adiabats as compared to the diabats.

The error bars in $\tau_{\pi\pi^*}$ and $\tau_{n\pi^*}$ give the error of the fit and does not include any other source of error; for example, it does not account for the degree of convergence of the wavepacket dynamics. In Figure S4, we show the calculated diabatic time constants versus the number of included initial conditions (1 to 16). This gives some indication of the degree of convergence of the wavepacket dynamics. In particular, the changes in the constants settle at around 5 to 10 initial conditions, exhibiting relatively small changes after this point (on the order of a few to around five fs).

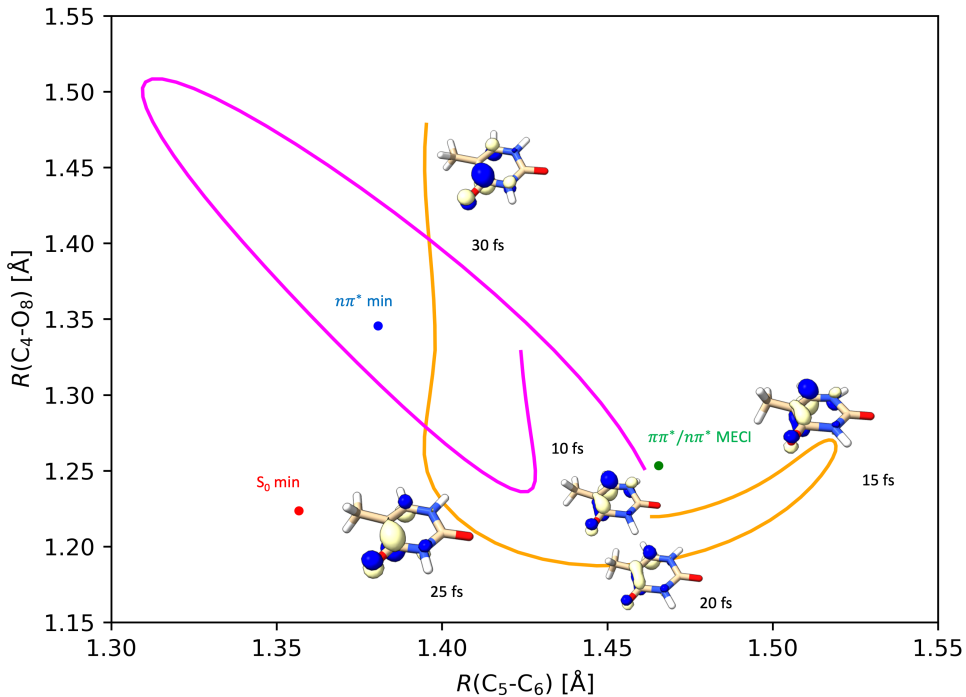

**Fig. S3**: A representative IC (IC 11) that features a population transfer from $S_2 - S_1$ retaining $\pi\pi^*$ character (orange) and a later transfer from $S_2 - S_1$ which changes from $\pi\pi^*$ to $n\pi^*$ (magenta). For the $S_1/\pi\pi^*$ trajectory, we visualize the dominant virtual NTO at intervals of 5 fs after spawn. Initially, the dominant virtual NTO is of $\pi\pi^*$ character, but as the trajectory nears the $n\pi^*$ minimum, it assumes $n\pi^*$ character.

## S1.3 Time constants observed in the simulated time-resolved oxygen-edge X-ray absorption spectrum

The feature at 526 eV is associated with the $n\pi^*$ state, as was previously indicated by calculations at the minimum on the $S_1(n\pi^*)$ surface [1]. In our simulated spectrum (Figure 2), we similarly find a feature at around 526 eV that grows in intensity in the first 50 fs to 70 fs. If the feature is in fact due to the $n\pi^*$ character of the wavepacket, as the evidence suggests, then we expect the $n\pi^*$ population and the intensity of the peak to be correlated.

To compare the peak intensity (at various times) and the diabatic $n\pi^*$ population, we shift the intensities such that the $n\pi^*$ intensity is zero at $t = 0$ and such that the maximum $n\pi^*$ intensity is equal to the maximum diabatic population of the $n\pi^*$ state. This shifted and rescaled intensity will then give a measure of the $n\pi^*$ diabatic population, provided the signal is caused by the $n\pi^*$ character of the wavepacket. Similarly, the $\pi\pi^*$ state has a feature at around 534 eV, which is masked by the ground state bleach in Figure 2. By the same argument, the intensity of this peak provides a measure of the population of the $\pi\pi^*$ state. We similarly rescale this $\pi\pi^*$ intensity so that the intensity is 1 at $t = 0$.

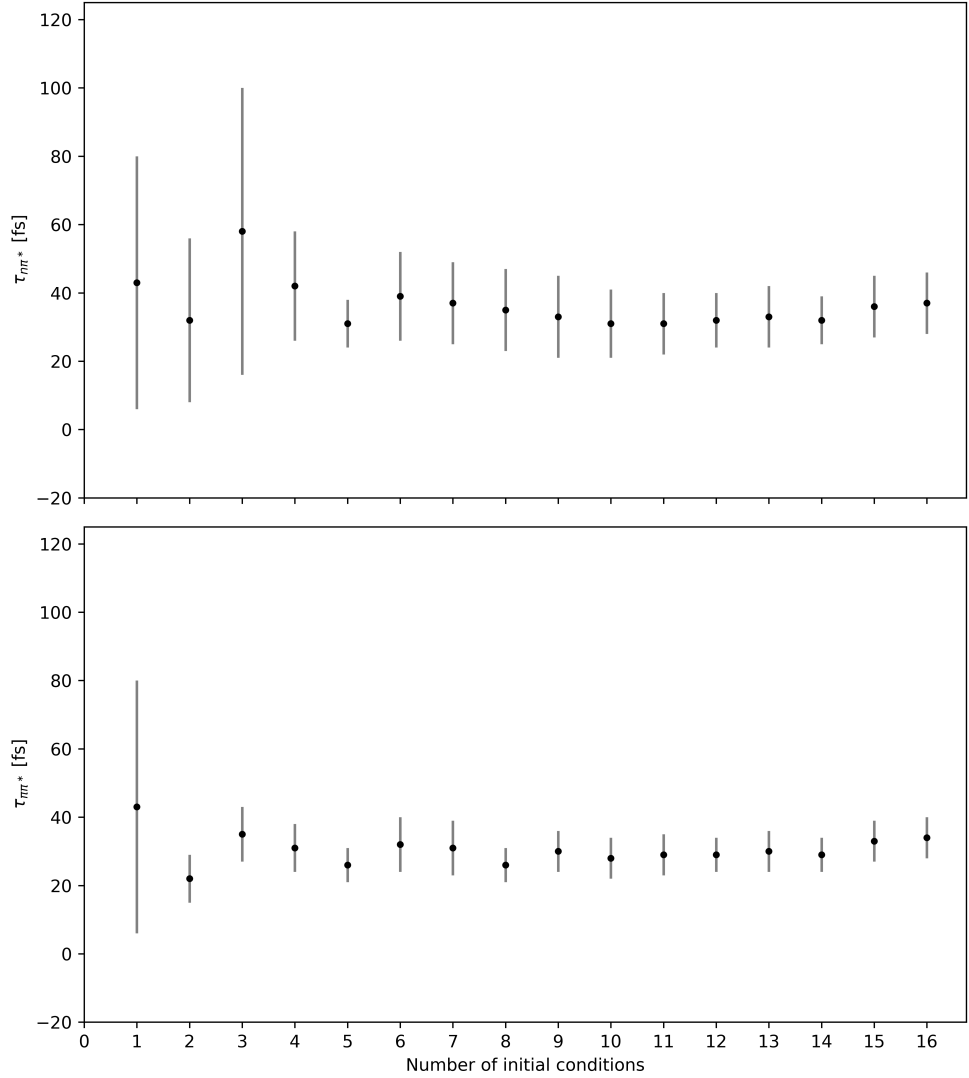

**Fig. S4**: Convergence of time constants calculated from diabatic populations. Upper panel: $\pi\pi^*/n\pi^*$ conversion. Lower panel: $\pi\pi^*$ decay. The error bars are given as two standard deviations.

The peak intensities, along with diabatic and adiabatic populations, are compared in Figure S5. We see that the diabatic population of the $n\pi^*$ state closely follows the growth of the 526 eV signal; similarly, the decay of the $\pi\pi^*$ diabat follows the decay of the 534 eV signal. The time constants are also in agreement with the diabats: we find $\tau_{n\pi^*} = 41 \pm 14$ fs (from the intensity) versus $\tau_{n\pi^*} = 37 \pm 9$ fs (diabat) and $\tau_{\pi\pi^*} = 27 \pm 4$ fs (intensity) versus $\tau_{\pi\pi^*} = 34 \pm 6$ fs (diabat).

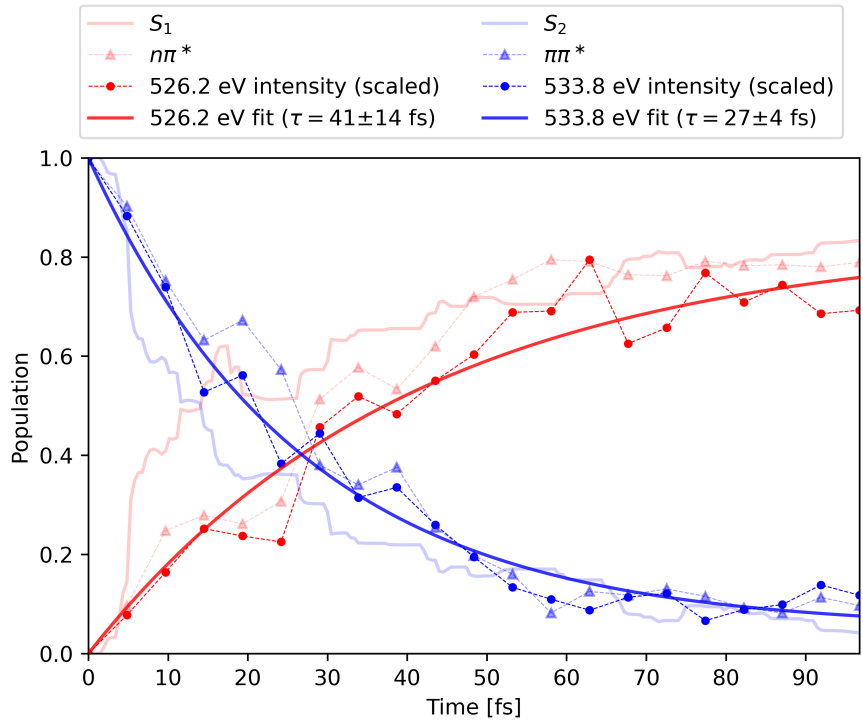

**Fig. S5**: Scaled intensities at 526.2 eV and 533.8 eV, as well as adiabatic populations for $S_1$ and $S_2$ and diabatic populations for $n\pi^*$ and $\pi\pi^*$ with associated lifetimes. The error bars on the characteristic time $\tau$ is reported as two standard deviations.

## S2 Simulated oxygen-edge X-ray absorption spectrum

The X-ray absorption of the excited states of thymine was calculated using a local development version of the eT program [2]. To compute the absorption strengths, we used CCSD/cc-pVDZ for the valence excited states and CC3/cc-pVDZ for the core excited states, where the core states were determined with the core-valence separation approximation [3]. Using CCSD for the valence states ensures that the valence excited states in the dynamics simulation match the valence excited states in the X-ray absorption calculations. All spectra were shifted by $-0.5$ eV to match the experimental ground state bleach at 531.4 eV [1].

In the first 800 au ($\sim$19.4 fs), we chose to calculate the absorption spectra at time steps of 20 au ($\sim$0.5 fs) to follow the decay of the $\pi\pi^*$ and the appearance of the $n\pi^*$ states more closely. From 800 au to 4000 au ($\sim$96.76 fs), a time step of 200 au ($\sim$4.84 fs) was chosen. To generate smooth spectra, individual excitations were broadened using Gaussian functions with full width at half maximum (FWHM) of 0.3 eV. For the false color plots the data was additionally interpolated along the time axis. This was done using Gaussian interpolation as provided by `matplotlib`'s [4] `imshow` or by broadening individual excitations with two dimensional Gaussians with FWHM 0.3 eV along the energy axis and FWHM 10 fs to 70 fs along the time axis.

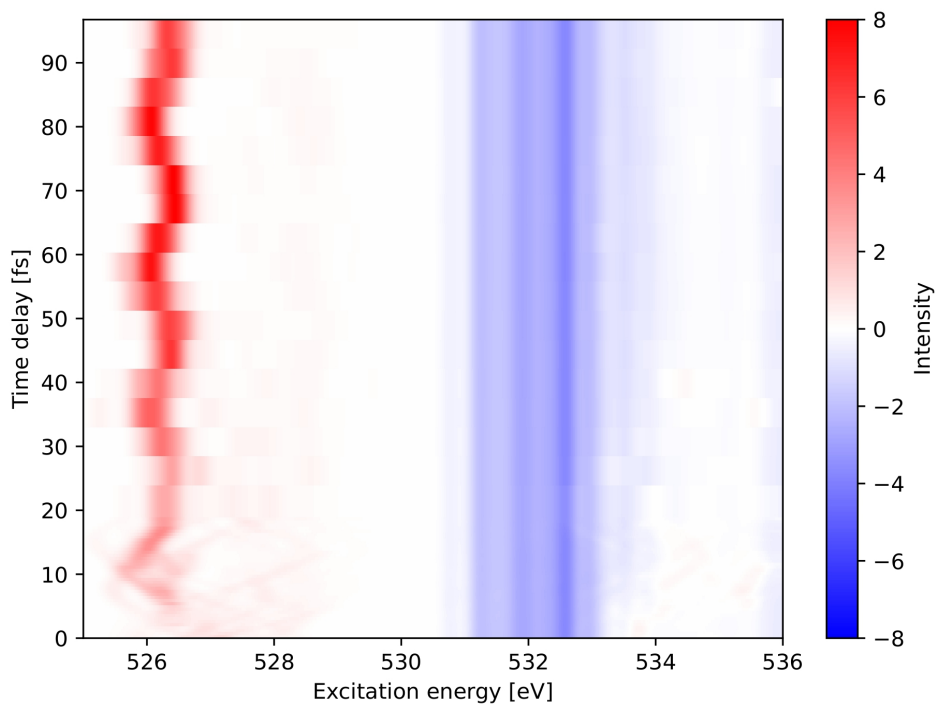

**Fig. S6**: False color plot without smoothing in the time direction.

## S3 Spawning geometries and the accessibility of the conical intersection seam along the $C_5$-$C_6$ stretching coordinate

In Figure S15, we plot the spawning geometries in terms of their $C_5$–$C_6$ and $C_4$–$O_8$ bond distances. Here, "spawning geometries" refers to geometries where new basis functions (TBFs) are created, which gives an indication of where we may see a significant transfer of population between the states. The distribution of spawning geometries shows that the seam is accessible along a $C_5$–$C_6$ lengthening coordinate that connects the Franck-Condon point ($S_0^{\min}$) and the minimum-energy conical intersection ($\pi\pi^*/n\pi^*_{\mathrm{MECI}}$). All but two of the spawns correspond to transfer from $S_2$ to $S_1$.

## S4 Wigner sampling: selection of initial conditions

The initial conditions (positions and momenta) used in the 16 AIMS runs were sampled from a harmonic Wigner distribution based on the ground state equilibrium geometry, determined at the CCSD/aug-cc-pVDZ level. The converged ground state

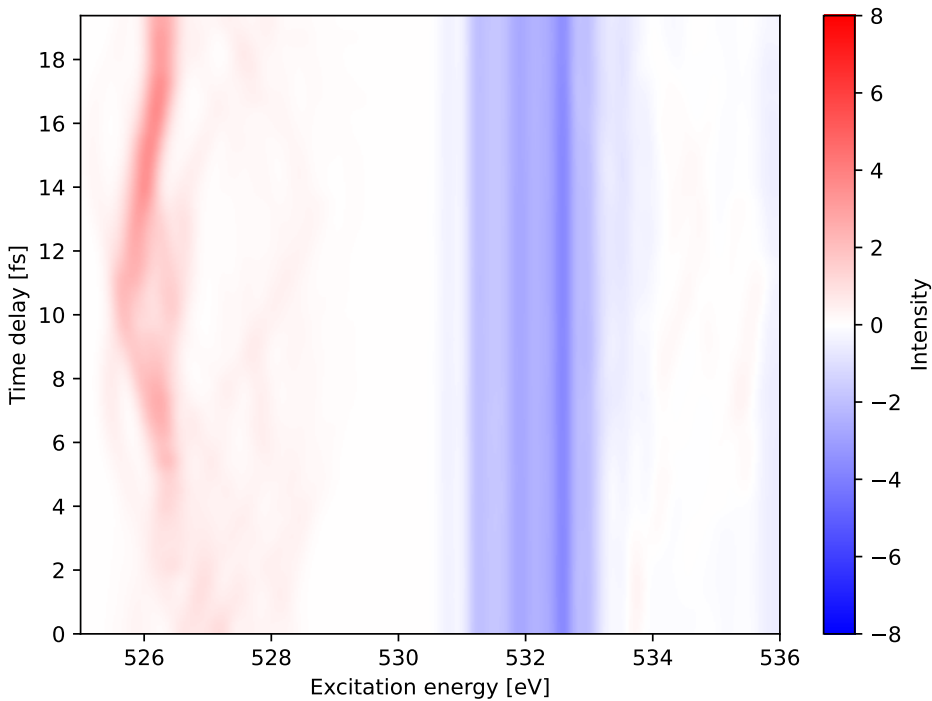

**Fig. S7**: Simulated thymine oxygen-edge X-ray absorption spectra for thymine for the first 800 au in time. Theoretical spectra are computed every 20 au and interpolated with Gaussians using `matplotlib`'s [4] `imshow`. For smoothing along the energy direction Gaussians with FWHM of 0.3 eV were used.

geometry, along with the harmonic frequencies, are given in Table S1. The harmonic frequencies were determined by evaluating the molecular Hessian, which was performed with $e^T$ using central differences ($\tau = 10^{-4}$ au) on the analytical CCSD gradients evaluated at displacements from the equilibrium geometry. The Hessian was used to generate the Wigner samples, which we have numbered as ICs 5–20. ICs 0–4 were used in initial tests to debug the interface and were not used in the dynamics simulations. For reference, ICs 6 and 12 lead to N-H dissociation, while 7 and 14 were re-run with SCCSD because they terminate in complex energy regions when run with CCSD. Positions and momenta of the individual samples are also available [5].

## S5 Stationary points on the excited state surfaces: the minimum energy $\pi\pi^*/n\pi^*$ conical intersection and the $n\pi^*$ minimum geometry

To interpret the wavepacket dynamics, we also located stationary points on the excited state surfaces. In Tables S2 and S3, we list the geometry for the $n\pi^*$ minimum and the minimum energy conical intersection between the $\pi\pi^*$ and $n\pi^*$ states, respectively.

These stationary points were determined at the same level as the dynamics (CCSD/cc-pVDZ).

**Table S1**: Ground state equilibrium geometry (in Å) and vibrational harmonic frequencies (in cm$^{-1}$) at the CCSD/aug-cc-pVDZ level.

| | | | | | | | | |
|---|---|---|---|---|---|---|---|---|
| C |  1.626856184467 | −0.090172437156 |  0.013282935761 | 106 | 536 |  906 | 1414 | 1812 |
| C | −0.197850164697 |  1.572388519191 | −0.033300408283 | 144 | 548 |  977 | 1424 | 3052 |
| C | −0.740709833470 | −0.739313238235 | −0.079521724593 | 174 | 607 | 1027 | 1449 | 3126 |
| C | −1.176243220430 |  0.598835299691 | −0.065959144242 | 276 | 680 | 1070 | 1469 | 3139 |
| C | −2.652953867469 |  0.918586250733 | −0.094848197888 | 292 | 739 | 1170 | 1494 | 3235 |
| N |  0.630784739942 | −1.058855452370 | −0.174427947644 | 392 | 763 | 1218 | 1524 | 3629 |
| N |  1.164641231138 |  1.205510451408 | −0.019672119665 | 392 | 780 | 1268 | 1739 | 3676 |
| O |  2.794813959974 | −0.394624876859 |  0.158035521208 | 460 | 811 | 1397 | 1783 | |
| O | −1.546761142159 | −1.810516595404 | −0.194583290615 | | | | | |
| H |  1.881303447604 |  1.906943526328 |  0.115985843483 | | | | | |
| H | −0.412611083215 |  2.640305248447 | −0.000321107921 | | | | | |
| H |  0.924394117785 | −1.993233441610 |  0.093943333488 | | | | | |
| H | −2.811463058403 |  2.006697839314 | −0.031363123742 | | | | | |
| H | −3.180294992549 |  0.445499759315 |  0.751943675010 | | | | | |
| H | −3.121321727307 |  0.554762000579 | −1.026259737912 | | | | | |

**Table S2**: $S_1(n\pi^*)$ minimum geometry (in Å) and vibrational harmonic frequencies (in cm$^{-1}$) at the CCSD/cc-pVDZ level.

| | | | | | | | | |
|---|---|---|---|---|---|---|---|---|
| C |  1.630701110258 | −0.071137104515 | −0.000217015039 |  90 | 460 |  801 | 1404 | 1869 |
| C | −0.209693539530 |  1.539822276693 |  0.000058364321 | 102 | 467 |  948 | 1428 | 3069 |
| C | −0.761020102709 | −0.825618775384 |  0.000090263142 | 104 | 498 | 1015 | 1446 | 3141 |
| C | −1.179422230038 |  0.591016860409 |  0.000089892193 | 215 | 527 | 1064 | 1482 | 3169 |
| C | −2.656435643450 |  0.893598512239 |  0.000225801361 | 249 | 582 | 1155 | 1494 | 3278 |
| N |  0.627861839137 | −1.031930751046 | −0.000192591426 | 257 | 605 | 1220 | 1499 | 3651 |
| N |  1.145275760481 |  1.226181730571 | −0.000168258378 | 340 | 752 | 1261 | 1544 | 3698 |
| O |  2.822138135024 | −0.341269451406 | −0.000301832749 | 396 | 760 | 1296 | 1633 | |
| O | −1.528917478290 | −1.778211641752 |  0.000340887262 | | | | | |
| H |  1.848723077280 |  1.951968102483 | −0.000249352114 | | | | | |
| H | −0.437664838893 |  2.607440739349 |  0.000178809529 | | | | | |
| H |  0.943714834308 | −1.996305266184 | −0.000110287090 | | | | | |
| H | −2.829665811383 |  1.981197603193 | −0.000029746644 | | | | | |
| H | −3.140793068686 |  0.454894321068 |  0.887377393240 | | | | | |
| H | −3.141002043508 |  0.454452844282 | −0.886592327609 | | | | | |

## S6 Natural transition orbitals for the initial conditions that end in N-H dissociation

Two of the initial conditions are found to lead to N-H dissociation via a channel that involves a $\pi\sigma^*$ state. In Figures S13 and S14 we track this process for one these (IC

Table S3: $S_1(n\pi^*)/S_2(\pi\pi^*)$ minimum energy conical intersection geometry (in Å) at the CCSD/cc-pVDZ level

| | | | |
|---|---:|---:|---:|
| C |  1.620884204303 | −0.107517382267 | −0.042077345164 |
| C | −0.175634683514 |  1.618168806981 | −0.079419662615 |
| C | −0.778255418212 | −0.815707290539 | −0.050927310541 |
| C | −1.173750723137 |  0.548693928425 | −0.059187433420 |
| C | −2.634883093476 |  0.878862929277 | −0.041363924381 |
| N |  0.652954617850 | −1.060239304783 | −0.061438342476 |
| N |  1.128940987061 |  1.228004642700 | −0.030710272580 |
| O |  2.828783658498 | −0.280358835324 | −0.019138916975 |
| O | −1.519845543610 | −1.823734952390 | −0.033898210074 |
| H |  1.880975271175 |  1.913878716948 |  0.019668160698 |
| H | −0.396377317533 |  2.680221014786 | −0.171050605554 |
| H |  0.948152824540 | −2.027583594230 | −0.121911512159 |
| H | −2.799214915932 |  1.965091087367 |  0.048680697186 |
| H | −3.133549718077 |  0.370648144289 |  0.805126231045 |
| H | −3.144410128255 |  0.521765471910 | −0.956125307028 |

6). The other (IC 12) shows similar behavior. Figure S14 shows the energy of the first three adiabatic states along the parent TBF for IC6. We see that $S_3$ becomes nearly degenerate with $S_2$, and this changes the character of the latter from $\pi\pi^*$ to $\pi\sigma^*$, as can be seen from the NTOs shown in Figure S13. Later, $S_2$ becomes nearly degenerate with $S_1$ and the $\pi\sigma^*$ character is again transferred, now to $S_1$. The majority of the IC's population (74%) is on the $\pi\sigma^*$ state when it dissociates. The same is true of the other IC (68%). Since the ICs are terminated when dissociation occurs, we cannot determine whether the remaining population would also eventually lead to N-H dissociation. Assuming that the only dissociative events are the ones observed, we find a lower estimate for dissociative events in the simulation of 9%. We should emphasize, however, that this should be taken as a rough estimate, given the number of samples on which the estimate is based (our simulation only includes 16 ICs). More samples are needed to provide an accurate estimate of how common this channel is in the AIMS/CCSD/cc-pVDZ dynamics.

In particular, we note that the sample used in the dynamics has a larger proportion of very short $N_1$-H distances than in a larger sample consisting of 10000 geometries. In the 16 ICs we sampled, 13% has an $N_1$-H bond length of less than 0.90 Å and 31% of less than 0.95 Å, compared to 3% and 12% found in the larger sample, respectively. This suggests that the rough estimate of 9% for the dissociative $\pi\sigma^*$ pathway may be an overestimate, although actual dynamics simulations need to be run to decide what the precise percentage is.

Among the initial conditions specifically selected to have a short N-H bond length (less than 0.90 Å), which we here label as 5, 8, 30, 96, 131, 150, 165, 167, 177, 189, 191, 193, 370, 411, 495, 512, 530, we find that ICs 131, 177 (both with $N_1$-H bond lengths greater than 2.0 Å), and 96 ($> 1.8$ Å), are dissociated and show $\pi\sigma^*$ character on a populated TBF. IC 5, 189, and 411 have $\pi\sigma^*$ character on a populated TBF with an $N_1$-H bond length of 1.5 Å, 1.274 Å, and 1.531 Å, respectively. In IC 512, one of the TBFs is dissociating ($N_1$-H bond length 1.753 Å) but does not have significant population. In several cases, these ICs encounter numerical issues and the simulation of these ICs have therefore not been completed. In the remaining ICs, we did not find

any involvement of the $\pi\sigma^*$ state. Figure S16 shows the plot in Figure 5D, where we have also included the ICs specifically chosen with short initial $N_1$-H. As in Figure 5D, we see that the dissociating ICs spawn at large $N_1$-H bond lengths.

## S7 Nitrogen-edge spectrum at Franck-Condon, at the $n\pi^*$, and at $Q = 1.0$ in the dissociative $N_1$-H pathway

In Figure S17, we give simulated nitrogen-edge X-ray absorption spectra at three geometries relevant to the dynamics: the ground state minimum, the $n\pi^*$ minimum, and a displaced geometry along the $N_1$-H dissociation pathway (in particular, $Q = 1.0$ for the normal mode associated with the bond stretch, corresponding to an $N_1$-H bond length of about 1.6 Å). Some observations can be made. First, the $n\pi^*$ and $\pi\pi^*$ signals are located at around 396–397 eV, with the $\pi\pi^*$ signal being the brighter of the two. We therefore expect a signal at 396–397 eV which mostly disappears in the first 50 fs as the wavepacket undergoes the $\pi\pi^*$ to $n\pi^*$ transition. At the extended $N_1$-H geometry ($Q = 1.0$), the $S_1(\pi\sigma^*)$ state has a signal that is shifted down to around 395 eV, although the exact location of this peak is expected to move as the hydrogen atom dissociates. Furthermore, a ground state peak is shifting down significantly, to around 398 eV, because the gap between the ground state and $\pi\sigma^*$ state is becoming progressively smaller as the $N_1$-H bond is extended. If some of the population transfers to the ground state, this peak could also be visible in a nitrogen-edge spectrum.

## S8 Interface for AIMS dynamics with CC theory

The AIMS simulations were performed with the FMS program developed in the Todd J. Martínez group. Since the CCSD and SCCSD nuclear gradients and nonadiabatic coupling elements were implemented in the $e^T$ program [2], we have written an interface that connects FMS with $e^T$. Over the course of the simulation, FMS requests information from the selected electronic structure program, which, in our case, is $e^T$. FMS requests the nuclear gradients and derivative coupling elements evaluated at the centers of TBFs and at the centroids between pairs of TBFs. This is all the information required to propagate the nuclear wavefunction in AIMS [7, 8].

To easily retrieve this information, we implemented a few objects/routines that allow FMS (and other programs) to use $e^T$ as a library. In particular: routines to set up and finalize $e^T$, and drivers to compute gradients and couplings. The changes to $e^T$ are planned to be made publicly available in an upcoming release of the program. On the FMS side, the interface template for electronic structure programs already exists and has been copied. In FMS, we have added a new model/electronic structure program ($e^T$) and implemented the routine that will be invoked for this new model. In particular, FMS calls this routine when it requires new electronic structure information, passing to it a trajectory object (including the geometry), a specification of the states $(i, j)$, and whether or not to compute the coupling between them. The routine then invokes an $e^T$ driver to compute gradients and couplings and populates the FMS trajectory with the required data.

We need to make sure that the coupling elements in FMS have a consistent phase. For this we use a simple orbital phasing procedure. The trajectory objects in FMS store the previous orbital coefficients and excited state amplitudes. In each new timestep, we pass this information on to $e^T$ where the current orbitals are phased. The current and previous orbitals are compared by evaluating the matrix

$$A_{pq}(\boldsymbol{x}) = \sum_{\alpha\beta} C_{\alpha p}(\boldsymbol{x}_0) S_{\alpha\beta}(\boldsymbol{x}) C_{\beta q}(\boldsymbol{x}), \quad (9)$$

where $\alpha, \beta$ denotes atomic orbital indices and $p, q$ denotes molecular orbital indices. The orbital coefficients are denoted $C_{\alpha p}$ and the atomic orbital overlap matrix as $C_{\alpha\beta}$. Finally, $\boldsymbol{x}_0$ refers to the previous geometry and $\boldsymbol{x}$ to the current geometry.

If an orbital flips sign from $\boldsymbol{x}_0$ to $\boldsymbol{x}$, the corresponding diagonal term of $\boldsymbol{A}$ will be approximately $-1$. If two orbitals change order energetically, then the corresponding 2-by-2 block will be dominated by the off-diagonal components. We first look for orbitals that change their order, assuming that the orbital has changed order energetically if

$$|A_{ij}| + |A_{ji}| > |A_{ii}| + |A_{jj}|. \quad (10)$$

After this procedure finishes, the diagonal components of $\boldsymbol{A}$ will have numbers close to 1 or $-1$ on the diagonal and small off-diagonal components. If the diagonal element is $-1$, we change the sign of the corresponding orbital. In this way, we obtain a consistent phase for the orbitals, which allows us to compare current and previous excited state amplitudes consistently by evaluating the state overlap matrix

$$S_{ij}(\boldsymbol{x}) = \boldsymbol{R}_i(\boldsymbol{x}_0)^T \boldsymbol{R}_j(\boldsymbol{x}). \quad (11)$$

If the diagonal is close to $-1$, we flip the sign of the excited state vector $\boldsymbol{R}_i(\boldsymbol{x})$. This should ensure a consistent phase for the coupling. The $\boldsymbol{S}$ matrix is also used to detect intersection jumps. If the off-diagonal elements dominate, where we again use the condition in Equation (10), then FMS will reject the timestep (provided that the state that flips is important for the dynamics) and attempt a shorter timestep.

## S9 Simulated UV-vis absorption spectrum

The UV-vis absorption spectrum was calculated at the CC3/aug-cc-pVDZ+KBJ(3-4) level for the ground state equilibrium geometry, determined with CCSD/aug-cc-pVDZ. To improve the description of possible Rydberg states in the spectrum, a set of Rydberg-type functions with quantum numbers $n = \{3, 3.5, 4\}$, generated according to the prescription of Kaufmann, Baumeister and Jungen [9], was added. These additional function are denoted "KBJ($n$=3–4)".

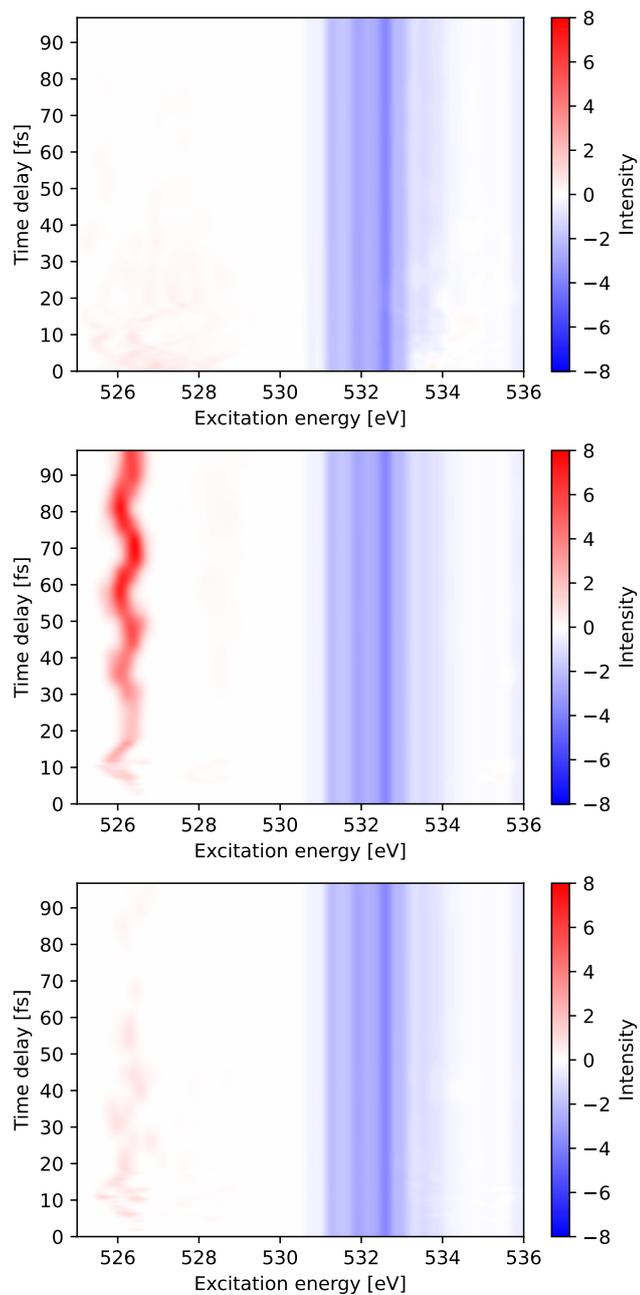

**Fig. S8**: Contribution from $\pi\pi^*$ (top) and $n\pi^*$ (center) states to the false color plot. The populated state was characterized by the sum of the oscillator strengths for the $S_0$ to $S_1$ and $S_0$ to $S_2$ transitions. $\pi\pi^*$ character was assumed if the sum was larger than 10 and and the populated state contributed more than 90% of the sum. $n\pi^*$ character if the populated state contributed 10% or less. Contributions that could neither be assigned to $\pi\pi^*$ nor $n\pi^*$ are plotted in the last panel.

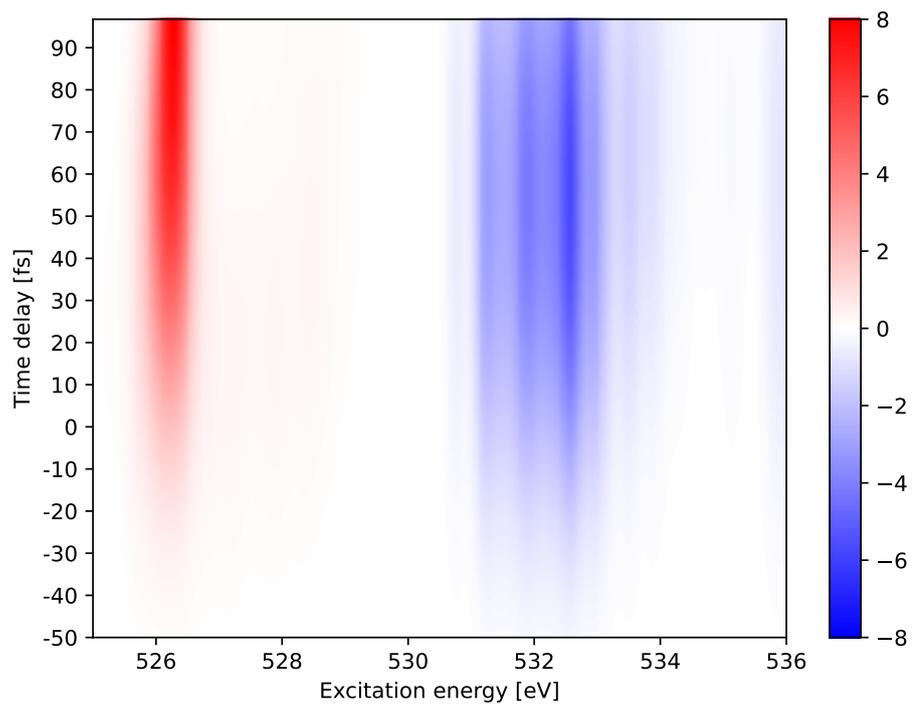

**Fig. S9**: False color plot with Gaussian smoothing using 70 fs FWHM and 0.3 eV FWHM. The absorption spectrum is computed in timesteps of 200 au (5 fs).

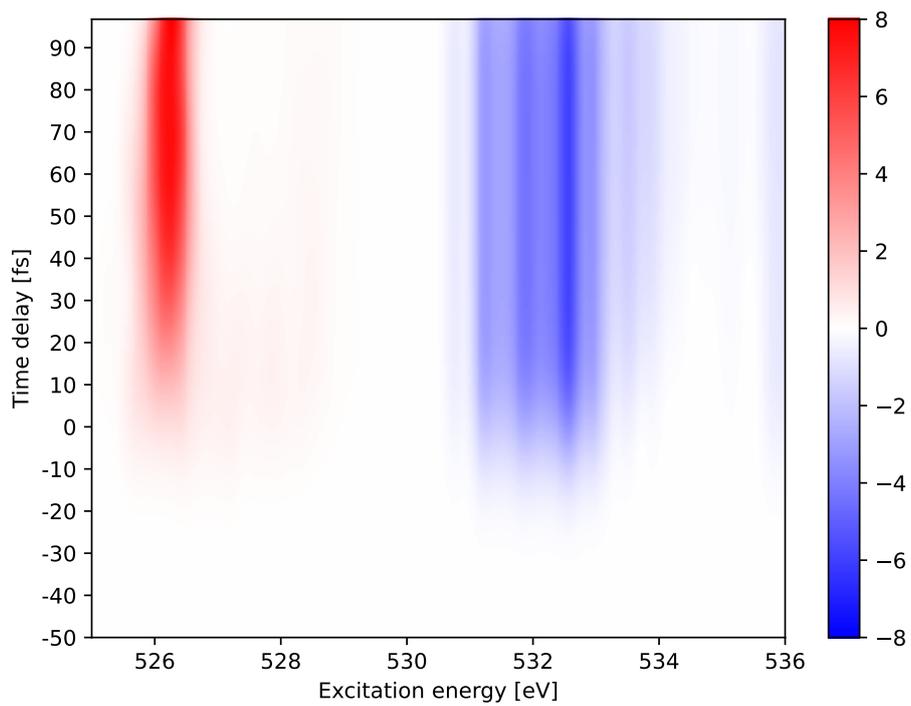

**Fig. S10**: False color plot with Gaussian smoothing using 30 fs FWHM and 0.3 eV FWHM. The absorption spectrum is computed in timesteps of 200 au (5 fs).

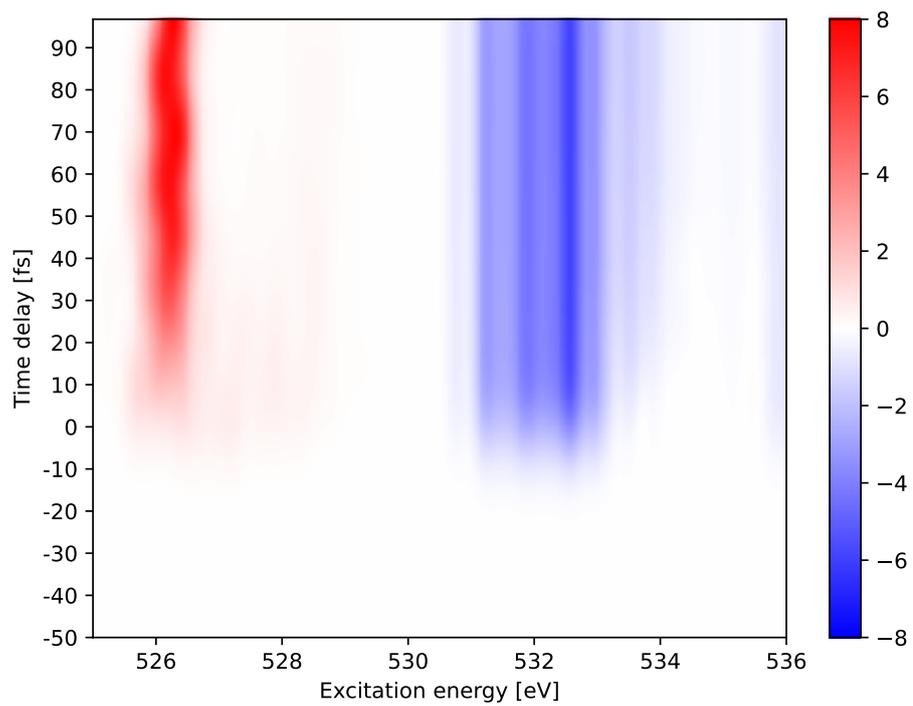

**Fig. S11**: False color plot with Gaussian smoothing using 20 fs FWHM and 0.3 eV FWHM. The absorption spectrum is computed in timesteps of 200 au (5 fs).

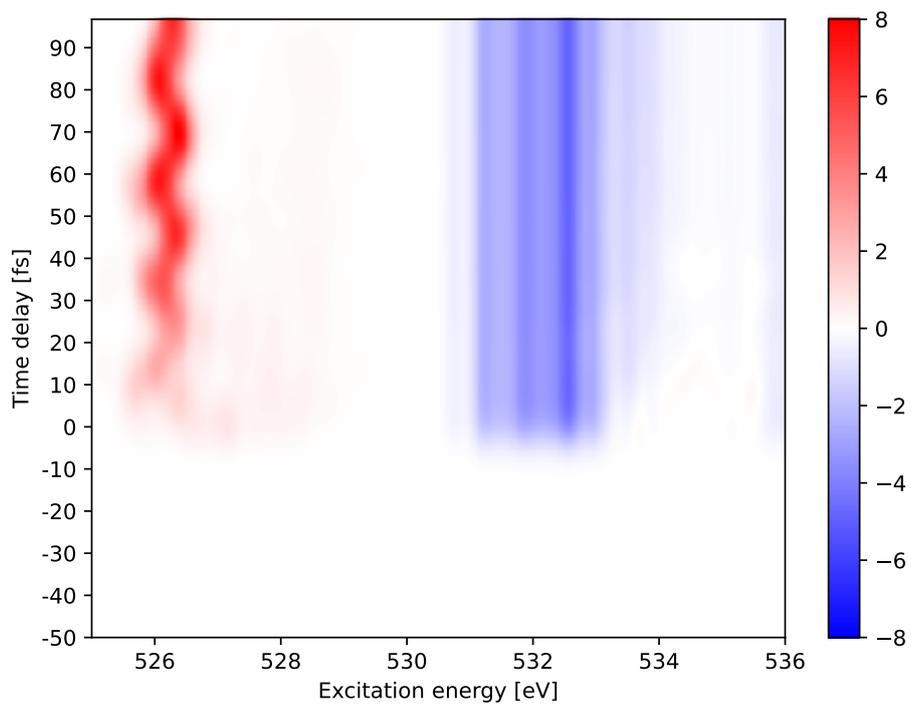

**Fig. S12**: False color plot with Gaussian smoothing using 10 fs FWHM and 0.3 eV FWHM. The absorption spectrum is computed in timesteps of 200 au (5 fs).

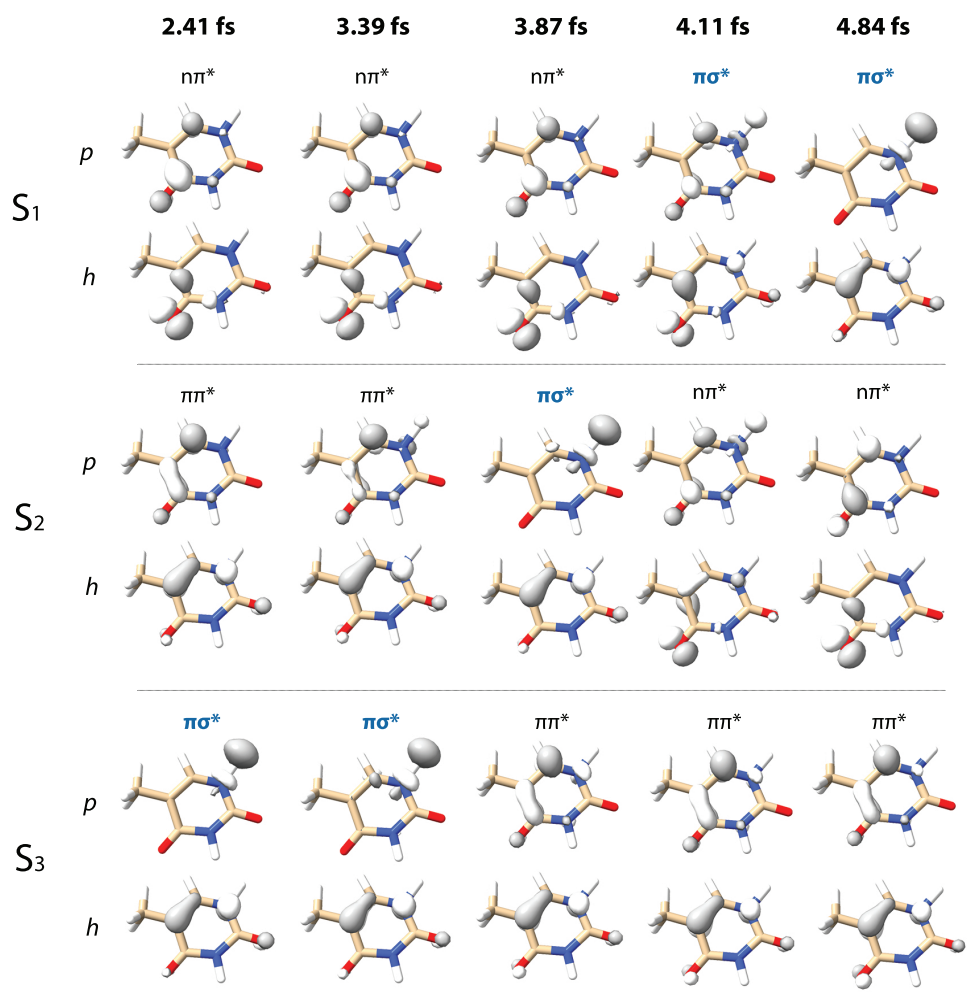

**Fig. S13**: Principal NTOs (particle denoted by $p$, hole by $h$) for $S_1$, $S_2$, and $S_3$ for IC 6, from 100 to 200 au. At 100 au $S_1$ has primarily $n\pi^*$ character, $S_2$ primarily $\pi\pi^*$, and $S_3$ is $\pi\sigma^*$. As the gap between $S_2$ and $S_3$ narrows, $S_2$ begins to assume more $\pi\sigma^*$ character, and by 160 au, the principal NTO of $S_2$ has $\pi\sigma^*$ character.

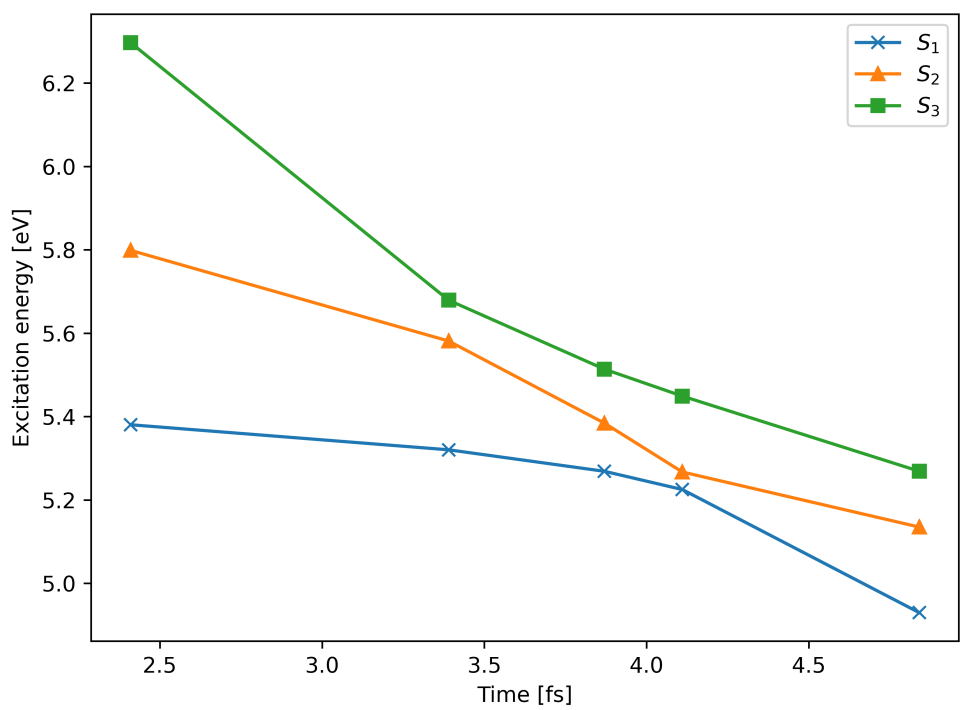

**Fig. S14**: Excitation energies for the first three excited states for IC 6 at 100, 140, 160, 170, and 200 au.

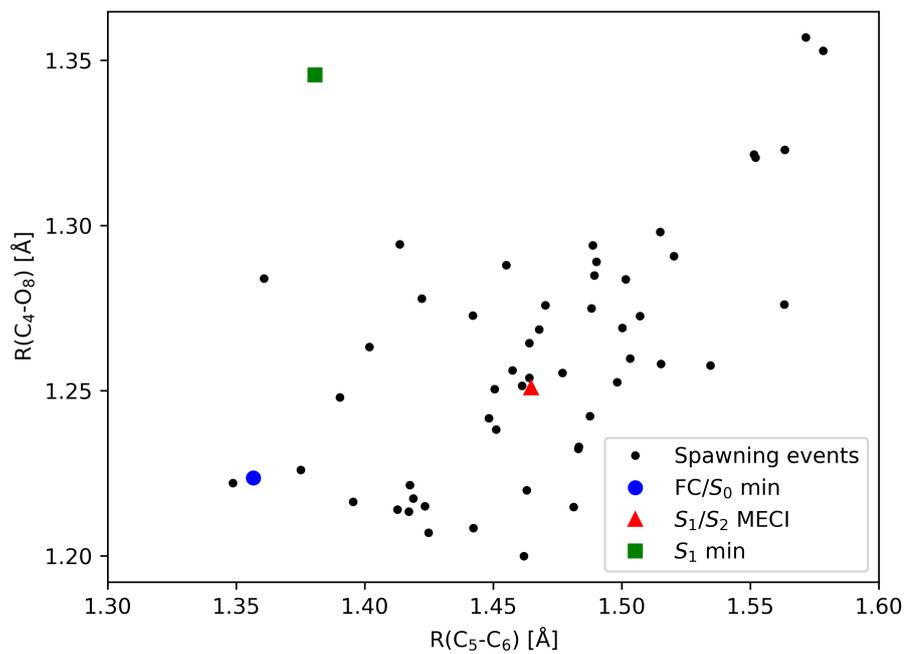

**Fig. S15**: Spawning geometries in the $C_5$–$C_6$ and $C_4$–$O_8$ bond coordinates. Black dots correspond to spawning geometries. The wavepacket accesses the $S_1/S_2$ MECI conical intersection seam primarily along a $C_5$-$C_6$ elongation from the initial Franck-Condon region, with a cluster of spawning events occurring close to the minimum energy conical intersection ($S_1/S_2$ MECI).

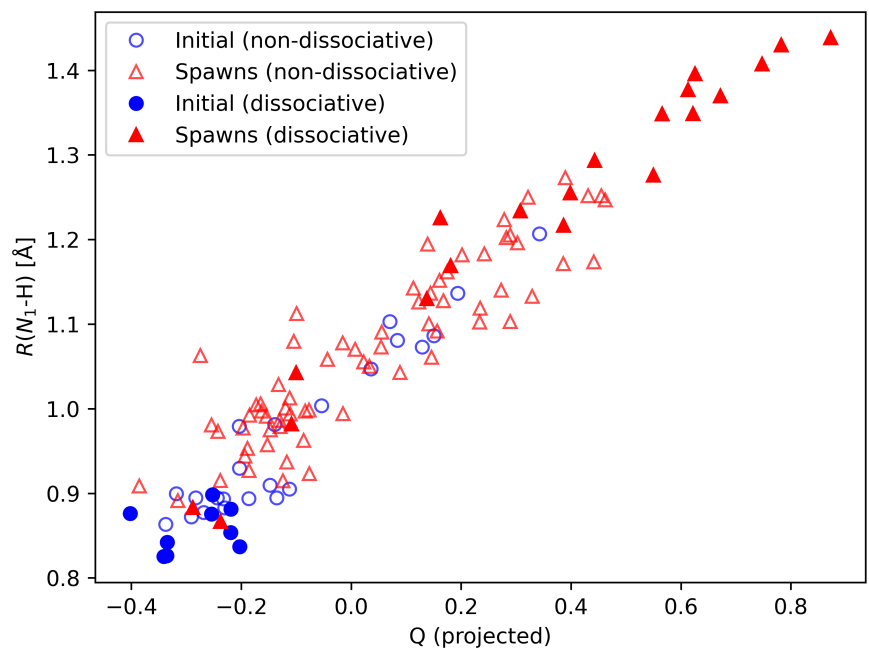

**Fig. S16**: Initial geometries and spawning geometries, in terms of the $N_1$-H bond length and the projected value for the $N_1$-H bond stretch normal mode $Q$, for the set of initial conditions used in the dynamics as well as the conditions specifically chosen to have short initial $N_1$-H bond lengths (less than 0.9 Å). Dissociating initial conditions appear to spawn at higher $N_1$-H bond lengths than non-dissociating conditions. This plot can be compared to Figure 5D, which only includes the initial conditions used in the simulation.

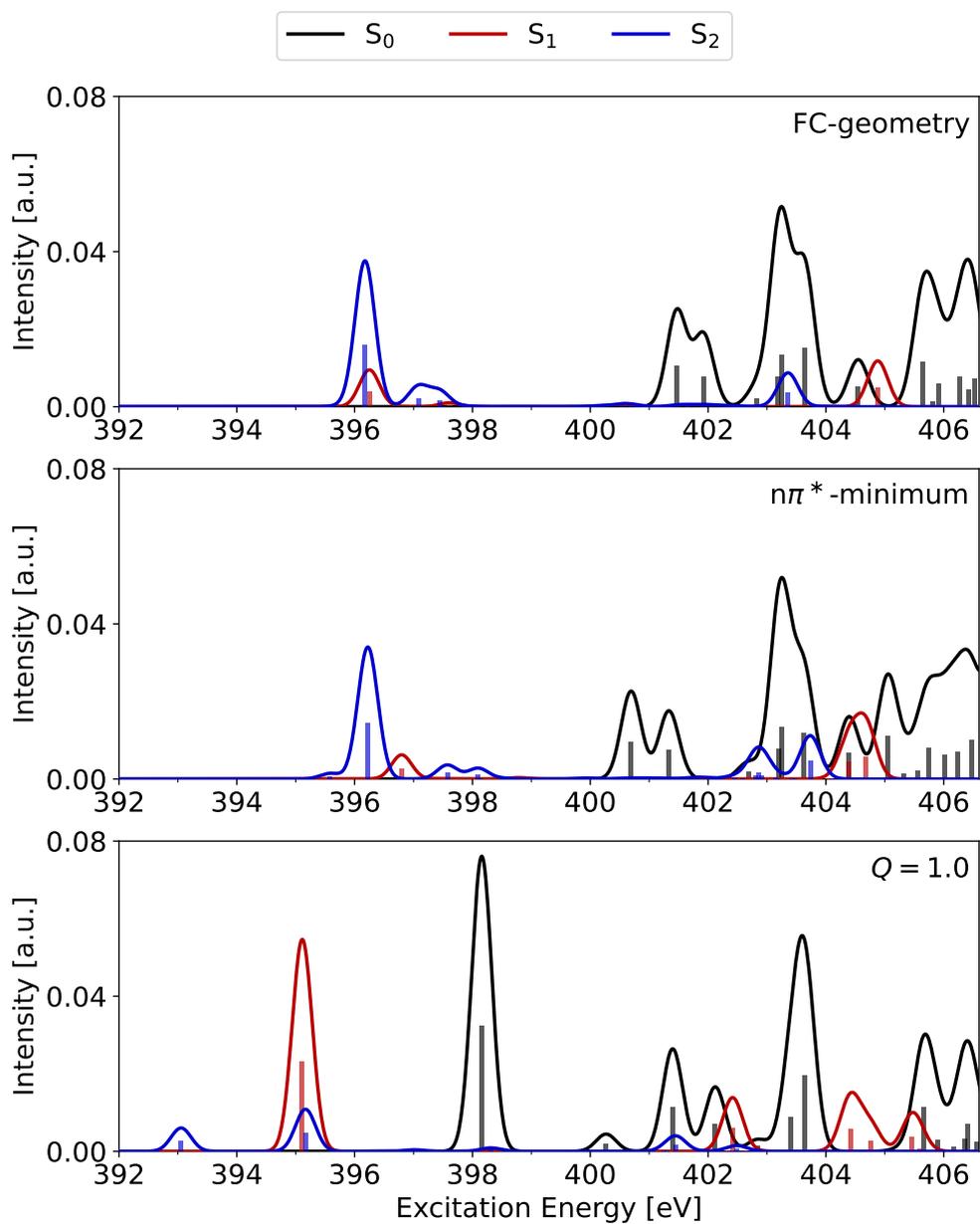

**Fig. S17**: Nitrogen-edge X-ray absorption spectrum at the Franck-Condon point, in the $n\pi^*$ minimum, and for a displacement $Q = 1.0$ along the $N_1$-H stretching mode. The spectra were calculated with CC3/cc-pVDZ for the core excited states and CCSD/cc-pVDZ for the ground state and the valence excited states. Individual excitations were broadened with Gaussian functions of 0.4 eV FWHM and all spectra were shifted by $-1.02$ eV to align with the experiment reported in Ref. 6.

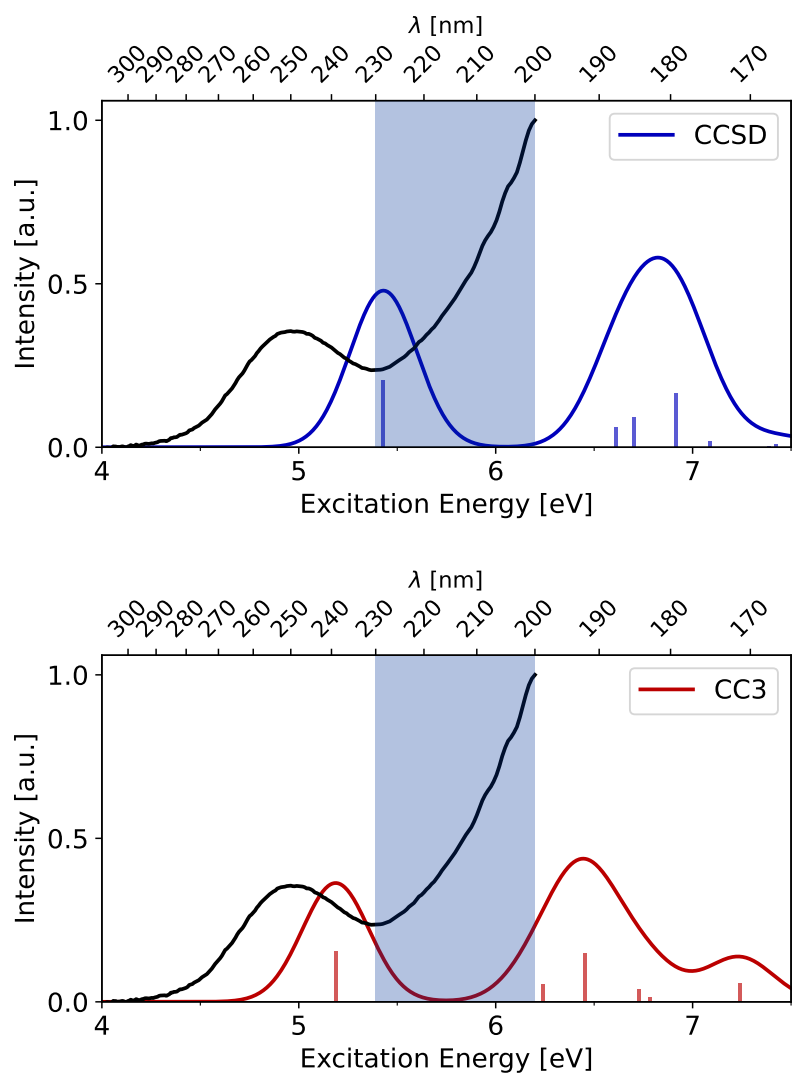

**Fig. S18**: Absorption spectrum of thymine calculated at the CCSD and CC3 level with aug-cc-pVDZ+KBJ(3-4) basis set at the ground state equilibrium geometry. The wavelength range between 230 nm and 200 nm is highlighted in blue. Individual excitations were broadened with Gaussian functions of 0.4 eV FWHM and no shift was applied.

**Table S4**: Excitation energies ($\omega$), oscillator strengths ($f$) and NTOs for the bright CC3/aug-cc-pVDZ+KBJ(3-4) excitations of thymine. An iso value of $0.03\,\text{electrons/bohr}^3$ was used.

| | $\omega$ [eV] | $f \times 100$ | hole NTO | particle NTO |
|---|---|---|---|---|
| $S_2$ | 5.19 | 15.47 | 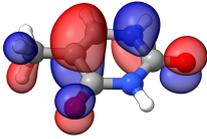 | 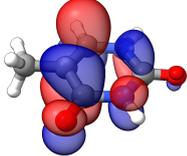 |
| $S_4$ | 6.24 | 5.20 | 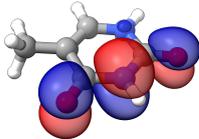 | 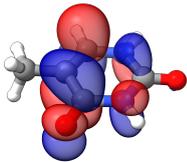 |
| $S_7$ | 6.46 | 14.91 | 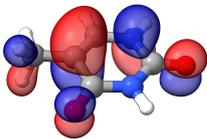 | 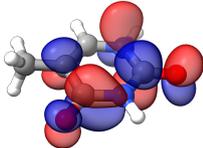 |
| $S_9$ | 6.73 | 3.91 | 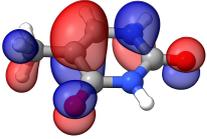 | 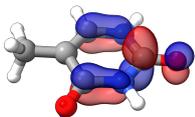 |
| $S_{11}$ | 6.78 | 1.43 | 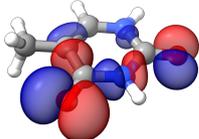 | 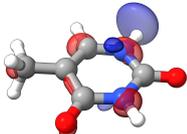 |
| $S_{15}$ | 7.24 | 5.75 | 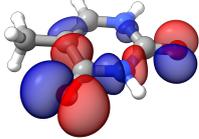 | 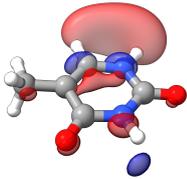 |



\end{document}